\documentclass[twocolumn]{aastex631}
\let\splitbox\relax
\usepackage{soul}

\usepackage{xcolor}

\usepackage{adjustbox}
\usepackage{amsmath}
\usepackage{appendix}

\begin{document}

\title{QPO signatures of disk restoration after Type--I X-ray bursts from 4U\,1636$-$536  }

\author{Suchismito Chattopadhyay}
\affiliation{Government Girls' General Degree College, 7, Mayurbhanj Road, Kolkata 700 023} 

\author{Jaiverdhan Chauhan}
\affiliation{School of Physics and Astronomy, University of Leicester, University Road, Leicester LE1 7RH, UK}
\affiliation{IUCAA, Pune University Campus, Ganeshkhind, Pune - 411007}

\author{Ranjeev Misra}
\affiliation{IUCAA, Pune University Campus, Ganeshkhind, Pune - 411007}

\author{Anne Lohfink}
\affiliation{Department of Physics, Montana State University, P.O. Box 173840, Bozeman, MT 59717-3840, USA}

\author{Rhaana Starling}
\affiliation{School of Physics and Astronomy, University of Leicester, University Road, Leicester LE1 7RH, UK}

\author{Priya Bharali}
\affiliation{Mahatma Gandhi Government Arts College, Mahe, Puducherry 673311, India}

\author{Soma Mandal}
\affiliation{Government Girls' General Degree College, 7, Mayurbhanj Road, Kolkata 700 023}

\begin{abstract}
Type--I thermonuclear bursts (TNBs) from neutron star low-mass X-ray binaries (NS LMXBs) originate on the neutron star's surface from the unstable burning of the accreted material. On the other hand, kHz quasi-periodic oscillations (QPOs) are thought to originate in the innermost regions of the in-spiralling accretion disk. {Type--I TNBs are expected to impact the inner accretion flow, and consequently the kHz QPOs, due to the intense radiation pressure}. In this work, we systematically study the evolution of the upper and the lower kHz QPOs immediately before and after a Type--I TNB on 4U\,1636$-$536 using {\em AstroSat} observations in the 3--20\,keV band. The analysis of the power-density-spectra show the presence of kHz QPOs within 200\,seconds before the onset of the Type--I burst. However, we have not detected any prominent signature of the same within 100--200\,sec after the burst. The kHz QPOs then re-emerges after $\approx$\,200\,sec. The fractional rms variation in the 3--20\,keV band drops by $\approx$\,5--6\,\%, supporting the non-existence of kHz QPOs in the 200\,sec post-Burst Zone. The time scale of 200\,sec coincides with the viscous time scale, highlighting a scenario where the inner disk is temporarily disrupted by the intense radiation from the Type--I TNB. The kHz QPO then re-establishes as the inner disk is restored.
\end{abstract}

\keywords {Unified Astronomy Thesaurus concepts: X-ray binary stars (1811); Accretion (14); Stellar accretion disks (1579)}

\section{Introduction} \label{sec:intro}
A low-mass X-ray binary (LMXB) consists of a compact object, either a neutron star (NS) or stellar-mass black hole (BH), that accretes matter from a less evolved companion star (which typically has a mass of $\lesssim 1 M_{\odot}$). Accretion occurs via Roche lobe overflow \citep{klitzing:frank1985accretion, Tauris2006}. These systems provide a unique laboratory to study accretion physics in strong gravity because they evolve on humanly observable time scales, as opposed to millions of years in active galactic nuclei \citep[weeks, months to years;][]{Martini2004, Hopkins2005, Corbel2013, Tetarenko2016}.

In particular, NSLMXBs show some extraordinary phenomena, including Type--I Thermonuclear X-ray bursts (TNBs), which appear as a sudden and dramatic increase in the X-ray brightness of accreting NSs, often reaching intensities many times greater than their persistent emission levels. In typical TNBs, the rising phase of X-ray flux ranges within $\approx$1 to 10 seconds before showing an exponential decay that spans tens to hundreds of seconds \citep[e.g.,][]{1993SSRv...62..223L,2003astro.ph..1544S,Galloway_2008, Galloway2021}. TNBs result from the unstable nuclear burning of accreted hydrogen and/or helium on the surface of the NS \citep{1993SSRv...62..223L,2003astro.ph..1544S,Galloway_2008}. Another kind of TNB can be observed in NSLMXBs called Type--II TNB, where the duration can vary in a wide range from a few milliseconds to a few fours. Type--II TNBs are thought to be arising due to accretion instabilities because the average power in Type--II TNBs is almost less by a factor of $\sim 10^2$ in comparison to the Type--I burst \citep{10.1093/mnras/stv330}. Various studies have shown evidence of interactions between the inner disk region and the burst emission \citep[e.g.,][]{Yu_1999, Zand2011, Keek2016, ZheYan2024}. Around a decade ago, \citet{refId0} discussed a possible interaction between kHz QPOs and Type--I TNBs in 4U\,1636$-$536 and 4U\,1608$-$52. The centroid frequency of the QPOs in NSLMXBs ranges from low frequency {{($\sim$ 5\,Hz--60\,Hz)}} to very high frequency ($\sim$\,1300\,Hz), showcasing a wide range of variability. The QPOs with low frequency (5--60\,Hz) and  Quality factor (Q) $>$ 2 are called Low-frequency QPOs (LFQPOs), and the QPOs with frequencies ranging mostly from 400--1300\,Hz are classified as High-Frequency QPOs or kHz QPOs \citep{kltizing:1989ARA&A..27..517V, klitzing:Van_Der_Klis_1997, klitzing:peille2015spectral, klitzing:wang2016brief}. These kHz oscillations often appear in pairs, with the higher centroid frequencies referred to as the upper kHz QPOs and the lower ones as the lower kHz QPOs. The time scale of these high-frequency variabilities is comparable to the dynamical time scale of the inner disk region. In their work, \citet{refId0} suggest that kHz QPOs may be affected by Type--I bursts on time scales ranging from a few tens to a few hundreds of seconds.  

{\emph{AstroSat}, with a high temporal resolution of $\sim$10,$\mu$s and a large effective area of $\sim$ 6000,cm$^2$ \citep{2017JApA...38...30A}, has detected kHz QPOs in several persistent atoll NSLMXBs, like 4U 1728$-$34, and 4U 1702$-$429, as reported in multiple studies \citep{Chauhan_2017, Anand_2024, Chattopadhyay_2024}.}




4U\,1636$-$536 is also a persistent atoll source, first reported by \citet{10.1093/mnras/169.1.7}, with an orbital period of 3.79\,hours, determined from optical light curve photometry \citep{Giles_2002}. Based on radius expansion, bursts reach the Eddington limit for hydrogen-rich (solar H fraction \(X \approx 0.7\)) and helium-rich (\(X = 0\)) material. \citet{Galloway_2006} estimated a distance to the source of $(6.0 \pm 0.5 \,\text{kpc}$) for a \(1.4 \, M_{\odot}\) neutron star, increasing to \(7.1 \, \text{kpc}\) for \(2 \, M_{\odot}\).  The system has a reported mass function of \(f(M) = 0.76 \pm 0.47 \, M_{\odot}\) and a donor mass ratio of \(M_2/M_{NS} \approx 0.21 - 0.34\), with an inclination angle of \(36^\circ\) -- \(74^\circ\) \citep{10.1111/j.1365-2966.2006.11106.x}. 4U\,1636$-$536 frequently exhibits burst oscillation in the Type--I TNBs, revealing a spin frequency of 290\,Hz and a first overtone near 580\,Hz \citep{1999ApJ...515L..77M}. {\em AstroSat} observations have also confirmed burst oscillations at \(\sim 581 \, \text{Hz}\) \citep{2021MNRAS.508.2123R}. Further, twin kHz QPOs have been detected in its power density spectrum using the Rossi X-ray Timing Explorer ({\it RXTE}), with lower kHz QPOs between 644--769\,Hz and upper kHz QPOs between \(\sim 900–1050 \,\text{Hz}\) \citep{1996ApJ...473L.135Z,2002MNRAS.336L...1J,10.1111/j.1365-2966.2005.09214.x,10.1111/j.1365-2966.2007.11943.x,2011ApJ...726...74L}. 


{In this work, we analyzed {\em AstroSat} data of 4U1636-52, with an emphasis to study the presence of kHz QPO before and after a TNB.} The paper is organized as follows: Section~\ref{sec:Observation Details} presents the observation details and basic data reduction procedures. Section~\ref{sec:Lightcurve} discusses the light curve analysis and the evolution of the source along the hardness-intensity diagram (HID). In Section \ref{subsec:temporal}, we explore the results of the advanced temporal analysis. Finally, Section~\ref{sec:discussion} concludes our findings and presents future implementations.


\section{Data reduction} \label{sec:Observation Details}
To date, {\em AstroSat} has observed 4U\,1636$-$536 on 11 occasions since 2016. Notably, Type--I TNBs have been detected in the light curves of the 10 observations of 4U\,1636$-$536. For this study, we selected four datasets where kHz QPOs were detected in addition to at least one Type--I TNB \citep{chattopadhyay2025spectraltiminganalysiskilohertzquasiperiodic}. Finally,  we select three of these four data sets (observation IDs G05\_195T01\_9000000530, hereafter observation 1, G05\_195T01\_9000000598 (observation 2), and G05\_002T01\_9000000616 (observation 3) for further analysis according to the criteria presented in Section \ref{sec:Lightcurve}. These three observations had an LAXPC exposure time of 39\,ksec, 38\,ksec, and 33\,ksec, respectively. The analysis of the LAXPC data was performed using {\tt LAXPCsoftware22Aug15}, which is publicly available on the ASSC website\footnote{\url{http://astrosat-ssc.iucaa.in/laxpcData}}. All orbit files were merged to generate a single level-2 event file. From this file, scientific products, including light curves, energy spectra, and power density spectra (PDS), were derived by applying good time intervals following the standard procedures outlined on the ASSC webpage\footnote{\url{http://astrosat-ssc.iucaa.in/uploads/threadsPageNew_SXT.html}}. It is important to note that all LAXPC PCUs—PCU 10, PCU 20, and PCU 30 data were utilized for detailed temporal analysis. 


\section{Lightcurve \& Hardness-Intensity Diagram} 
\label{sec:Lightcurve}

\begin{figure}
	\centering
		\includegraphics[width=0.48\textwidth]{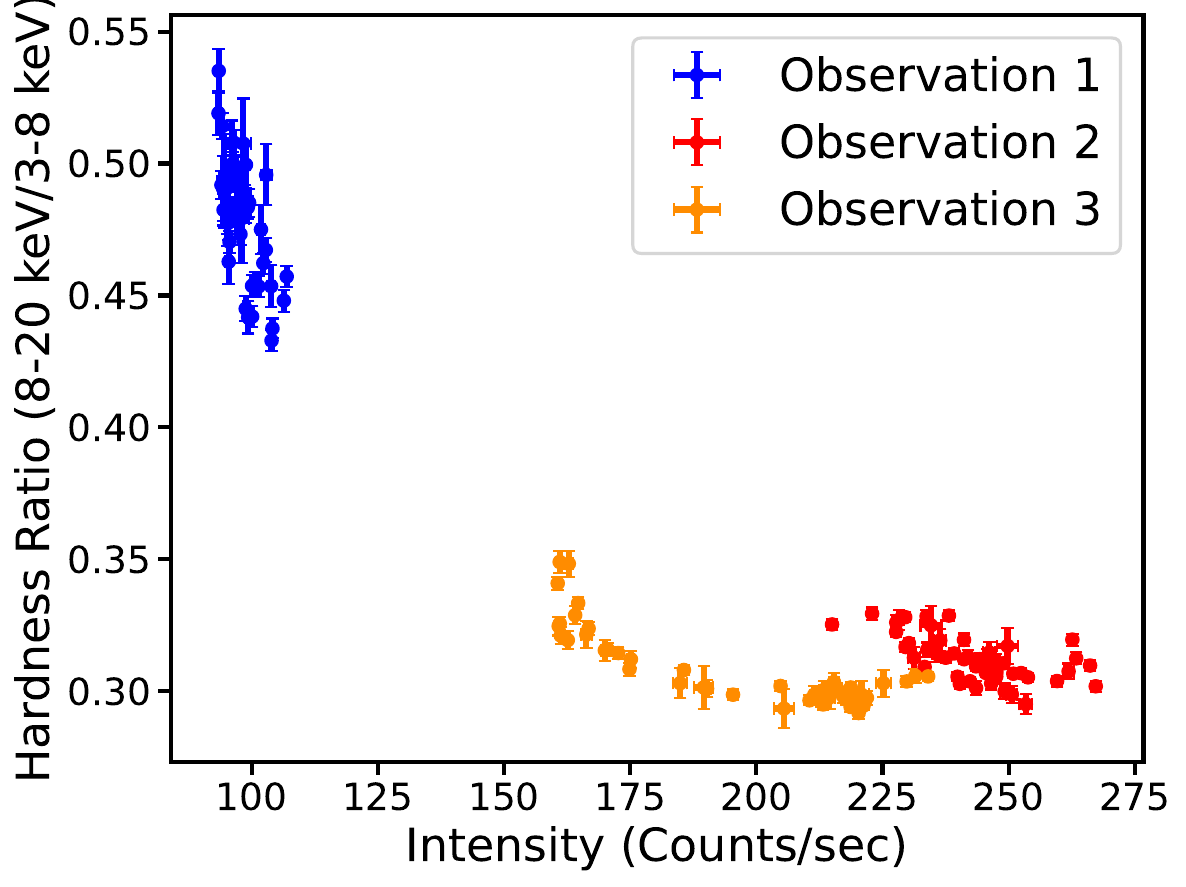}
	\caption{Hardness vs Intensity diagram using LAXPC PCU 20 to see the evolution of the source. The bin time is 1024\,sec. }
	\label{Fig00}
\end{figure}


\begin{figure}
	\centering
	\begin{tabular}{c}
		\includegraphics[width=0.48\textwidth]{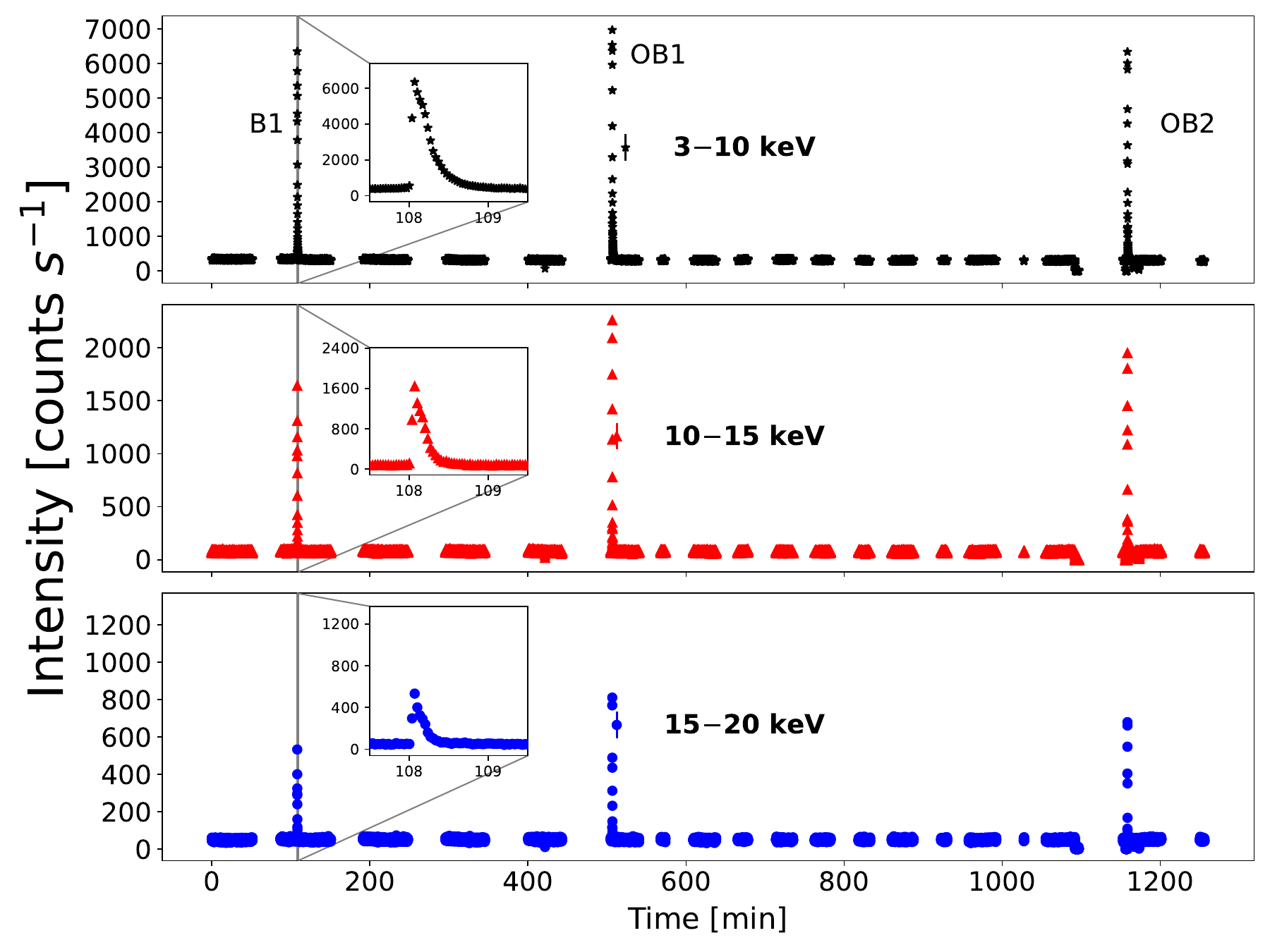} \\
		\includegraphics[width=0.48\textwidth]{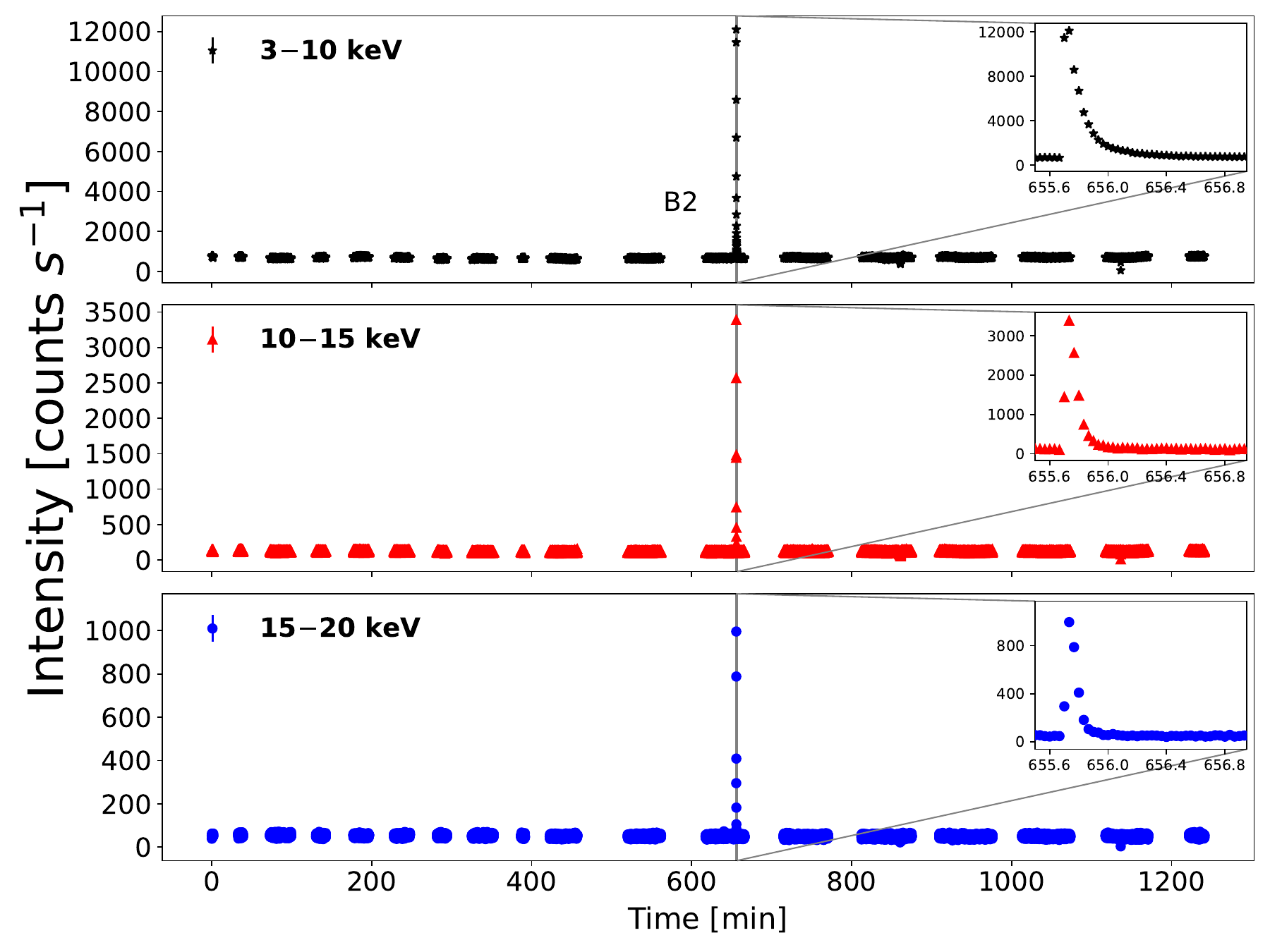} \\
		\includegraphics[width=0.49\textwidth]{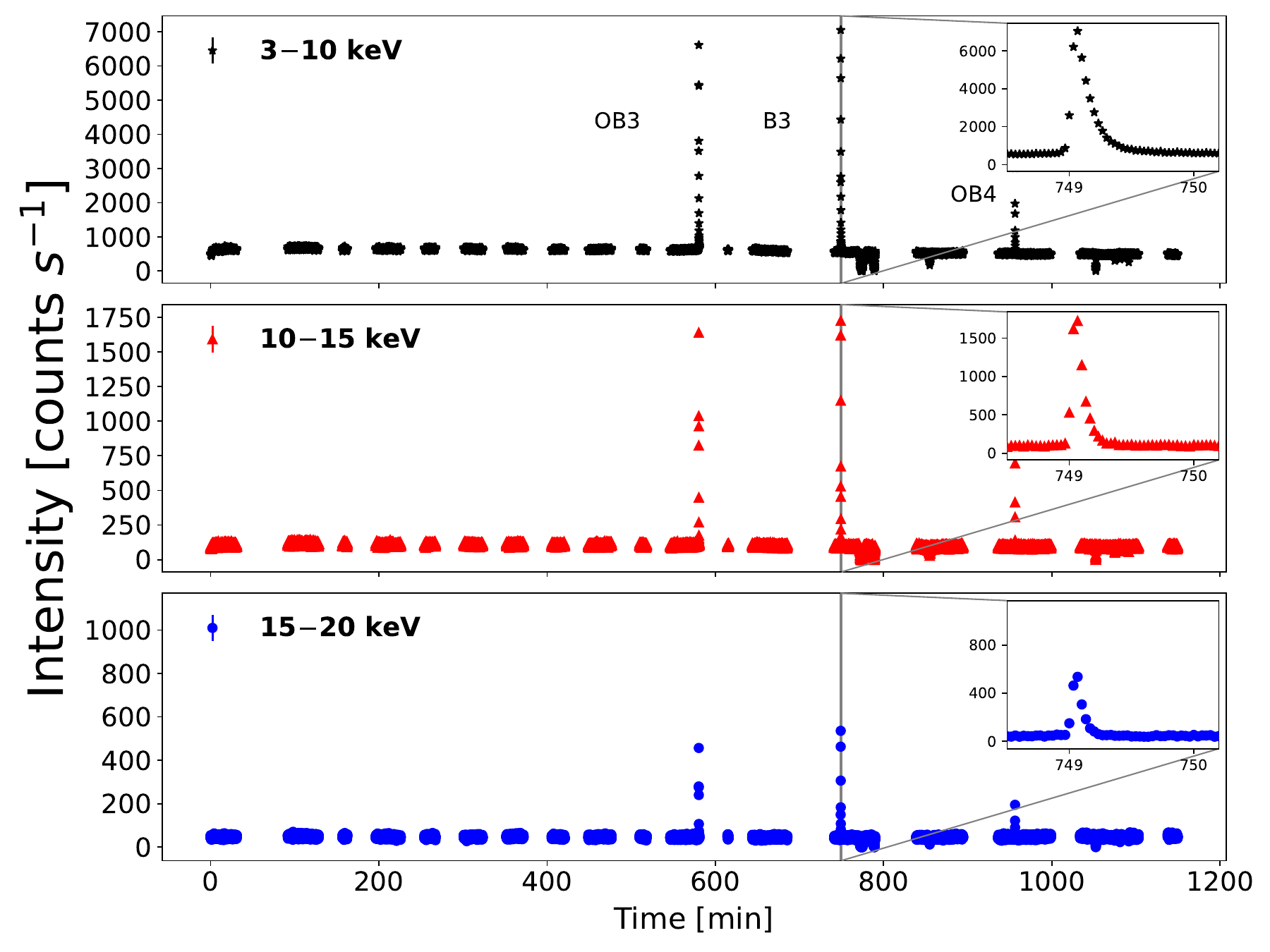} \\
	\end{tabular}
\caption{Light curves of observations 1, 2, and 3 in the 3--10, 10--15, and 15--20\,keV energy ranges. The abbreviation OB stands for the Omitted Burst, representing burst emissions that have not been considered for further analysis. B1 (TNB--1), B2 (TNB--2), and B3 (TNB--3) represent bursts considered for further analysis.}
\label{Fig01}	
\end{figure}


\begin{table*}
\centering
\normalsize  
\renewcommand{\arraystretch}{1.2}
\setlength{\tabcolsep}{3pt}

\caption{The best fitted parameters obtained from the fitting of the PDS using Lorentzian functions. ``Lorentzian 1" and ``Lorentzian 2" represent two Lorentzian components. $\nu$ is the centroid frequency, $\sigma$ is the FWHM. {SNR is calculated using: $0.5 \times (\mathrm{rms})^2 \times C_t \times \sqrt{T/2\sigma}$ \citep{2004astro.ph.10551V,2008MNRAS.384.1519B}, where $C_t$ is the total count rate, $T$ is the segment exposure, and rms is back scaled: $rms_{\mathrm{backscaled}} = (C_s / C_t) \times rms$, with $C_t = C_s + C_b$. $C_b$ is the count rate of the background, which remains more or less constant across the different segments of the different observations. For Obs 1 this is around $\sim152$\,counts\,sec$^{-1}$. For Obs 2 and Obs 3 these are $\sim156$ and $\sim145$\,counts\,s$^{-1}$, respectively.}}
\resizebox{\textwidth}{!}{%
\begin{tabular}{cccccccccc}
\hline
\hline
Obs & Comp & Param & \textbf{Pre-Burst} & \textbf{Post-Burst} & \textbf{Post-Burst} & \textbf{Post-Burst} & \textbf{Post-Burst} & \textbf{Post-Burst} & \textbf{Post-Burst} \\
 &  &  & \textbf{-200--0} & \textbf{0--200} & \textbf{200--400} & \textbf{400--600} & \textbf{600--800} & \textbf{800--1200} & \textbf{1200--1400} \\
\hline
1 & Lorentzian 1 & $\nu_1$ & 708 $\pm$ 13 & 708$^{\dagger}$ & 708$^{\dagger}$ & 719 $\pm$ 15 & 619 $\pm$ 19 & 623 $\pm$ 13 & - \\
  &        & $\sigma_1$ & 42 $\pm$ 20 & 42$^{\dagger}$ & 42$^{\dagger}$ & $< 61$ & $<25$ & 30 $\pm$ 14 & - \\
  &        & rms (\%) & 20 $\pm$ 3 & $<8.36$ & $<13$ & 17 $\pm$ 4 & $<$ 14 & 15 $\pm$ 3 & - \\
  &        & {SNR} & {6.7} & {1.1} & {2.6} & {3.6} & {3.90} & {4.0} & - \\
  & Lorentzian 2 & $\nu_2$ & - & - & - & - & 785 $\pm$ 33 & 811 $\pm$ 13 & 1130 $\pm$ 21 \\ 
  &        & $\sigma_2$ & - & - & - & - & $<25$ & 26 $\pm$ 12 & $<54$ \\
  &        & rms (\%) & - & - & - & - & $<15$ & 14.8 $\pm$ 3 & 15 $\pm$ 4.4 \\
  &        & {SNR} & - & - & - & - & {4.45} & {4.3} & {3.0} \\
  &        & Source CR ($C_s$) & 321 $\pm$ 25 & 313 $\pm$ 22 & 307 $\pm$ 22 & 303 $\pm$ 22 & 300 $\pm$ 22 & 295 $\pm$ 21 & 287 $\pm$ 20 \\
  &        & $\chi^2$/dof & 0.93/28 & 1.11/30 & 0.88/30 & 0.65/28 & 0.73/25 & 1.1/25 & 0.78/28 \\
\hline
Obs & Comp & Param & \textbf{Pre-Burst} & \textbf{Post-Burst} & \textbf{Post-Burst} & \textbf{Post-Burst} & \textbf{Post-Burst} & \textbf{Post-Burst} & \textbf{Post-Burst} \\
 &  &  & \textbf{-200--0} & \textbf{0--200} & \textbf{200--400} & \textbf{400--600} & \textbf{600--800} & \textbf{800--1200} & \textbf{1200--1400} \\
\hline
2 & Lorentzian 1 & $\nu_1$ & 738 $\pm$ 14 & 738$^{\dagger}$ & - & - & - & - & - \\   
  &        & $\sigma_1$ & $< 54$ & 54$^{\dagger}$ & - & - & - & - & - \\   
  &        & rms (\%) & 11 $\pm$ 4 & $<4.0$ & - & - & - & - & - \\
  &        & {SNR} & {3.33} & {0.5} & - & - & - & - & - \\
  & Lorentzian 2 & $\nu_2$ & - & - & 989 $\pm$ 21 & 1073 $\pm$ 20 & 1163 $\pm$ 10 & - & - \\  
  &        & $\sigma_2$ & - & - & 55 $\pm$ 35 & $37 \pm 26$ & $<30$ & - & - \\   
  &        & rms (\%) & - & - & 13 $\pm$ 3.5 & 10 $\pm$ 4 & 11 $\pm$ 4 & - & - \\
  &        & {SNR} & - & - & {4.41} & {3.1} & {4.3} & - & - \\
  &        & Source CR ($C_s$) & 699 $\pm$ 31 & 687 $\pm$ 27 & 674 $\pm$ 27 & 671 $\pm$ 27 & 677 $\pm$ 34 & - & - \\
  &        & $\chi^2$/dof & 1.26/28 & 1.21/30 & 0.72/28 & 0.74/28 & 0.73/28 & - & - \\
\hline  
Obs & Comp & Param & \textbf{Pre-Burst} & \textbf{Post-Burst} & \textbf{Post-Burst} & \textbf{Post-Burst} & \textbf{Post-Burst} & \textbf{Post-Burst} & \textbf{Post-Burst} \\
 &  &  & \textbf{-200--0} & \textbf{0--200} & \textbf{200--400} & \textbf{400--600} & \textbf{600--800} & \textbf{800--1200} & \textbf{1200--1400} \\
\hline
3 & Lorentzian 1 & $\nu_1$ & 698 $\pm$ 6 & 698$^{\dagger}$ & 648 $\pm$ 10 & 655 $\pm$ 55 & 739 $\pm$ 44 & 695 $\pm$ 10 & - \\   
  &        & $\sigma_1$ & $<13$ & 13$^{\dagger}$ & $<15$ & $<19$ & $<90$ & $<35$ & - \\
  &        & rms (\%) & 11 $\pm$ 3 & $<5.0$ & 10.8 $\pm$ 3 & 7 $\pm$ 1 & $<12$ & 14 $\pm$ 2 & - \\
  &        & {SNR} & {7.5} & {1.54} & {6.58} & {3.0} & {3.25} & {7.13} & - \\
  & Lorentzian 2 & $\nu_2$ & - & - & 968 $\pm$ 54 & 812 $\pm$ 40 & 988 $\pm$ 27 & 1015 $\pm$ 8 & - \\   
  &        & $\sigma_2$ & - & - & $<37$ & $<20$ & $<66$ & $<35$ & - \\   
  &        & rms (\%) & - & - & $<11$ & $<5$ & $<12$ & $<12$ & - \\
  &        & {SNR} & - & - & {4.2} & {1.74} & {3.8} & {5.3} & - \\
  &        & Source CR ($C_s$) & 562 $\pm$ 28 & 558 $\pm$ 15 & 540 $\pm$ 24 & 538 $\pm$ 24 & 542 $\pm$ 26 & 543 $\pm$ 24 & - \\
  &        & $\chi^2$/dof & 1.3/28 & 0.83/30 & 0.81/25 & 0.82/25 & 1.09/25 & 1.2/25 & - \\
\hline  

\end{tabular}
}
\label{Table3}
\end{table*}


\begin{figure*}
\begin{tabular}{c c}
Pre -200\,--\,0\,sec & Post 0\,--\,200\,sec\\
\includegraphics[width=0.45\textwidth]{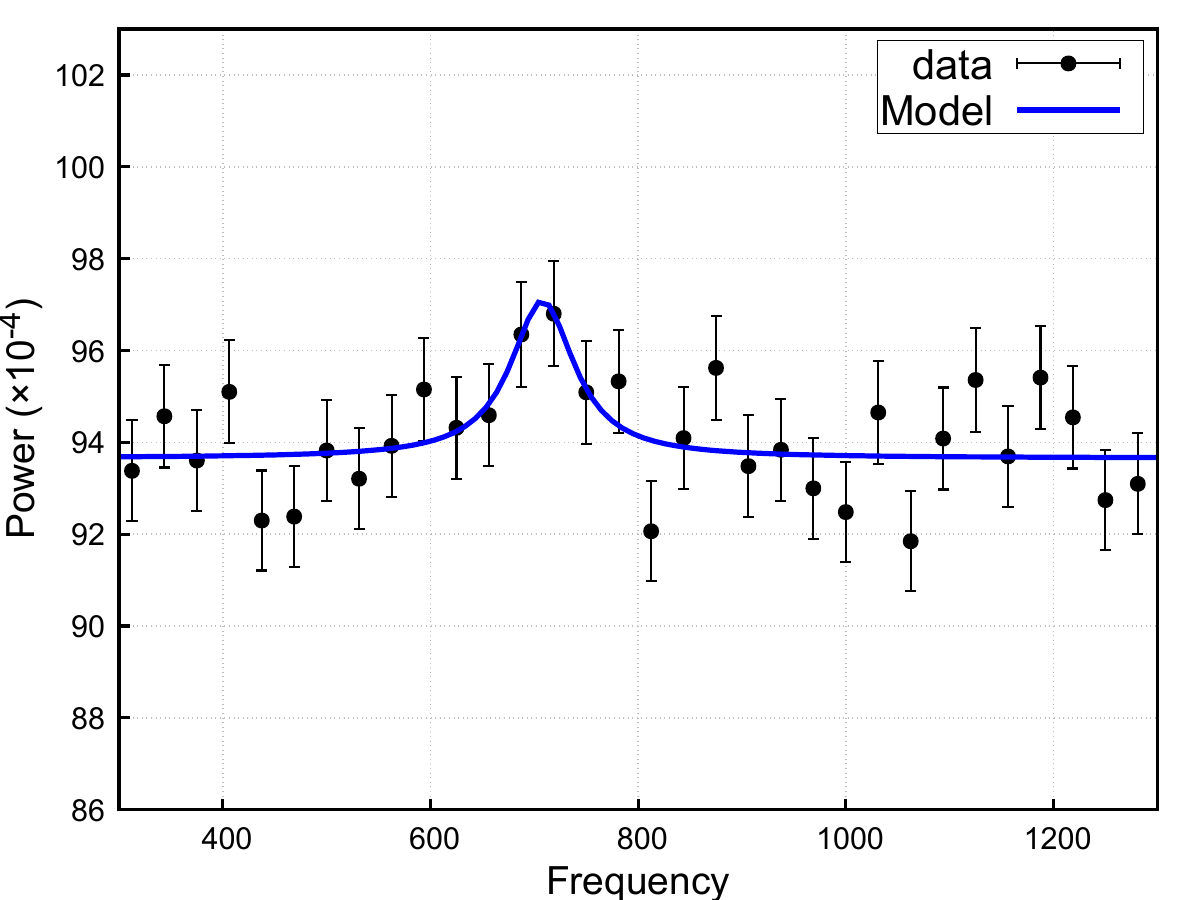} & \includegraphics[width=0.45\textwidth]{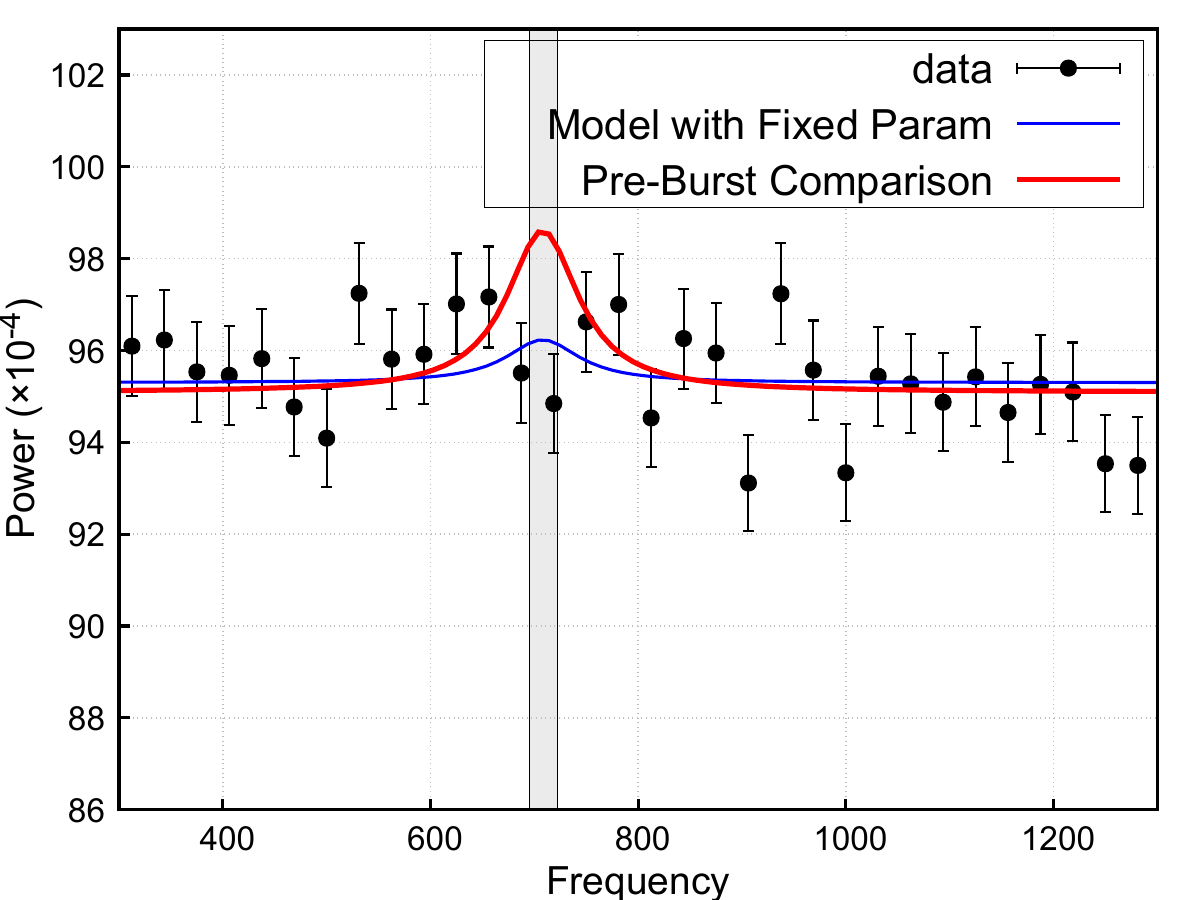}\\
Post 200\,--\,400\,sec & Post 400\,--\,600\,sec (*)\\
\includegraphics[width=0.45\textwidth]{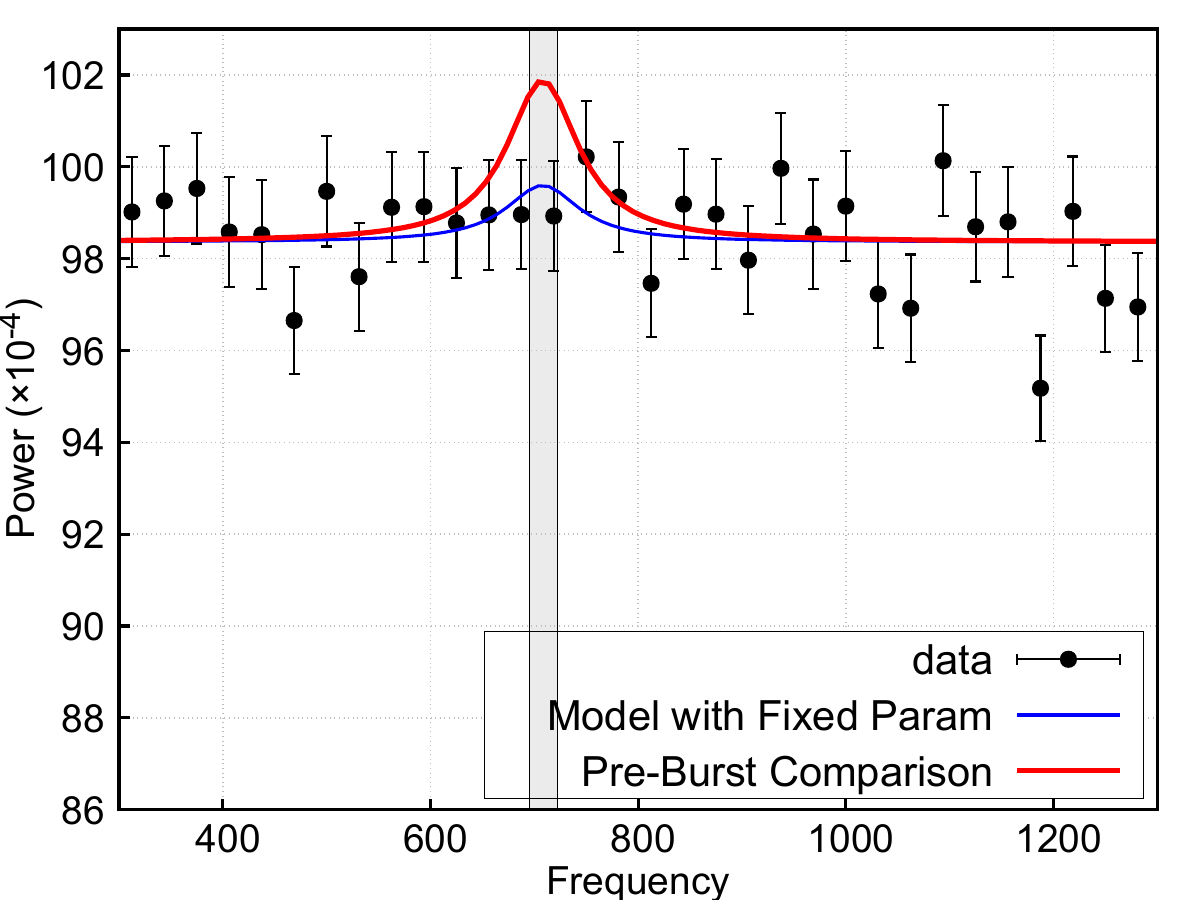} & \includegraphics[width=0.45\textwidth]{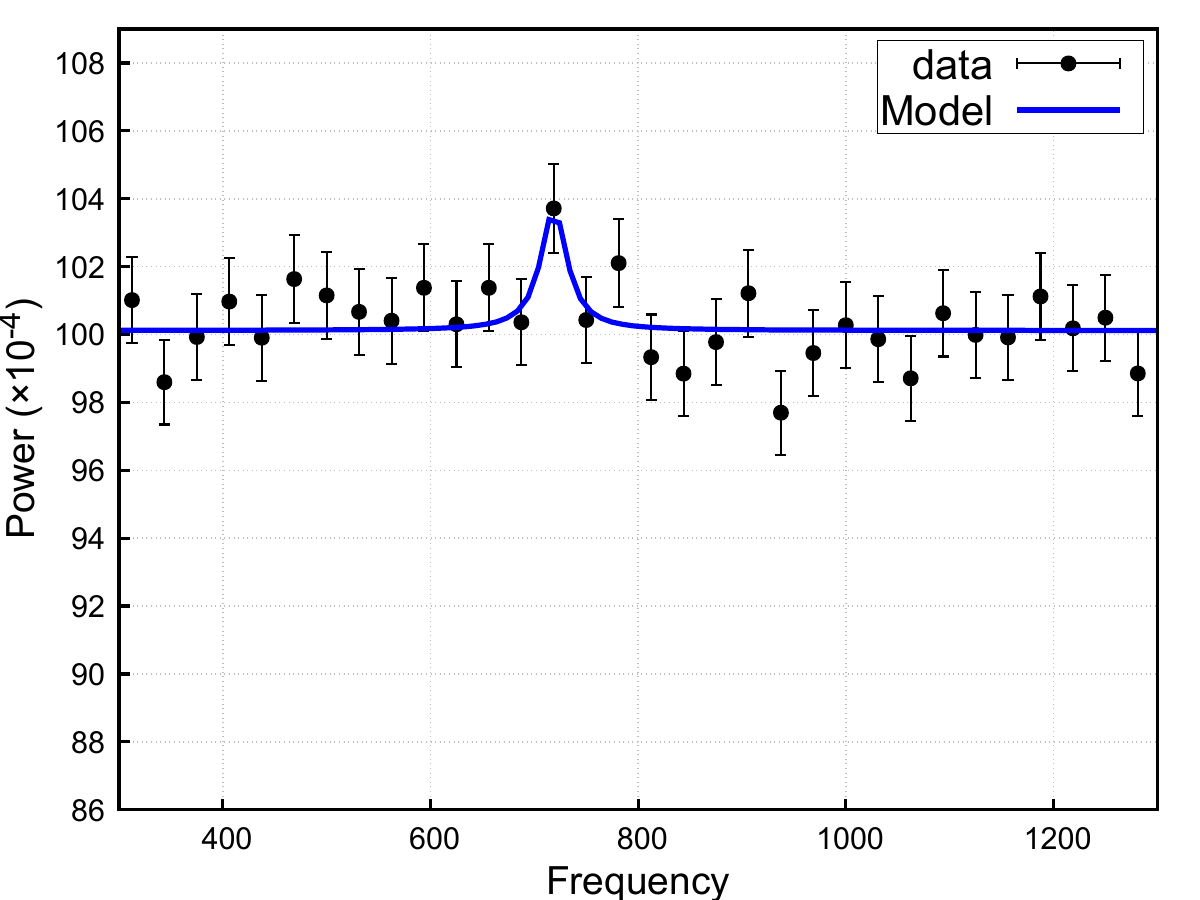}\\
Post 600\,--\,800\,sec (*) & Post 800\,--\,1200\,sec (*)\\
\includegraphics[width=0.45\textwidth]{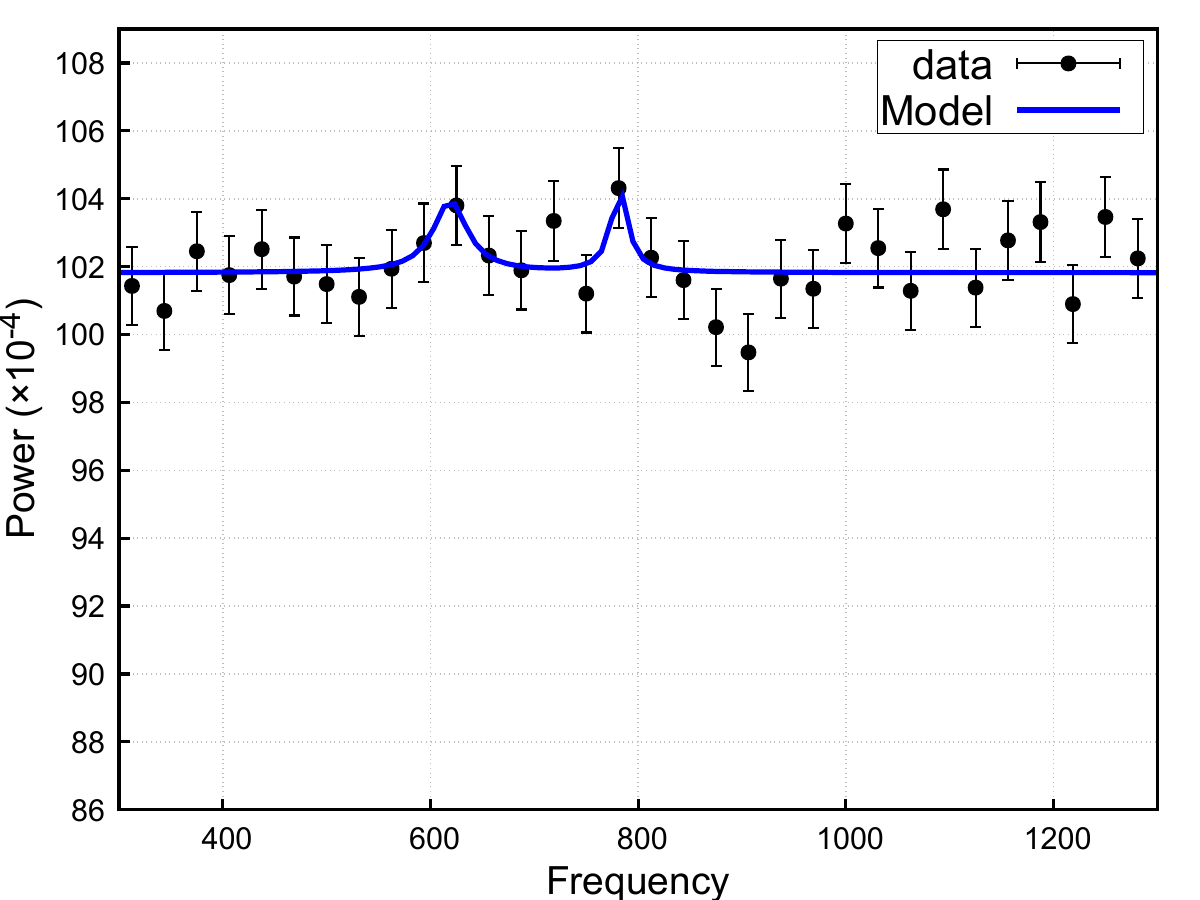} & \includegraphics[width=0.45\textwidth]{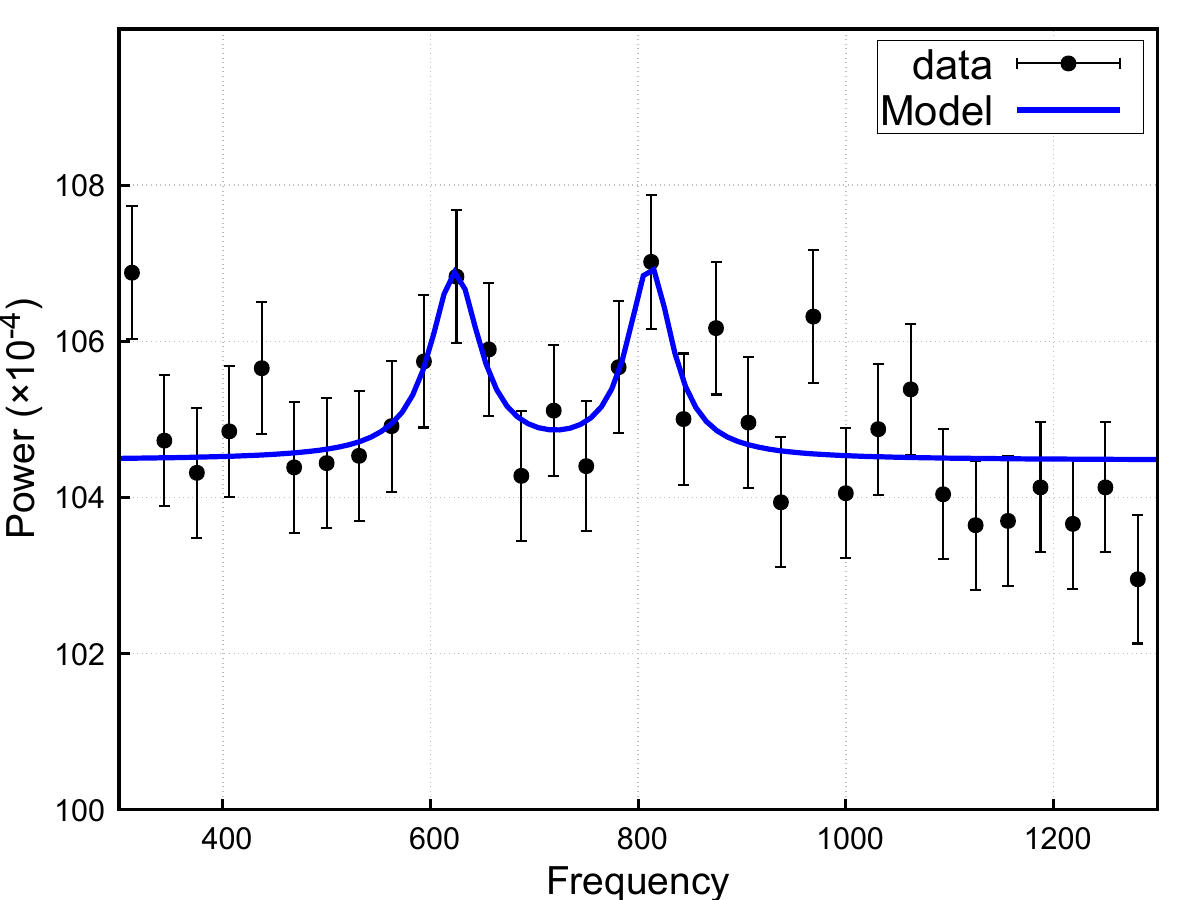}\\
\end{tabular}
\caption{The evolution of the kHz QPO in the pre-burst and the post-burst zone of the Type--I TNB of Observation 1. While the kHz QPO is present in pre-burst data, the same QPO is not present in the post-burst within 0--400\,sec data (we over-plot the pre-burst QPO with the blue line, highlighting the lack of signal). In some scenarios { (400--600\,sec, 600--800\,sec, 800--1200\,sec)}, we have changed the y-axis limit to show the prominent presence of the QPO, which has been marked with the *. The frequency is measured in Hz, and the power is normalized by rms.}
\label{Obs1-evol}
\end{figure*}


\begin{figure*}
\begin{tabular}{c c}
Pre -100\,--\,0\,sec (*) & Post 0\,--\,100\,sec\\
\includegraphics[width=0.45\textwidth]{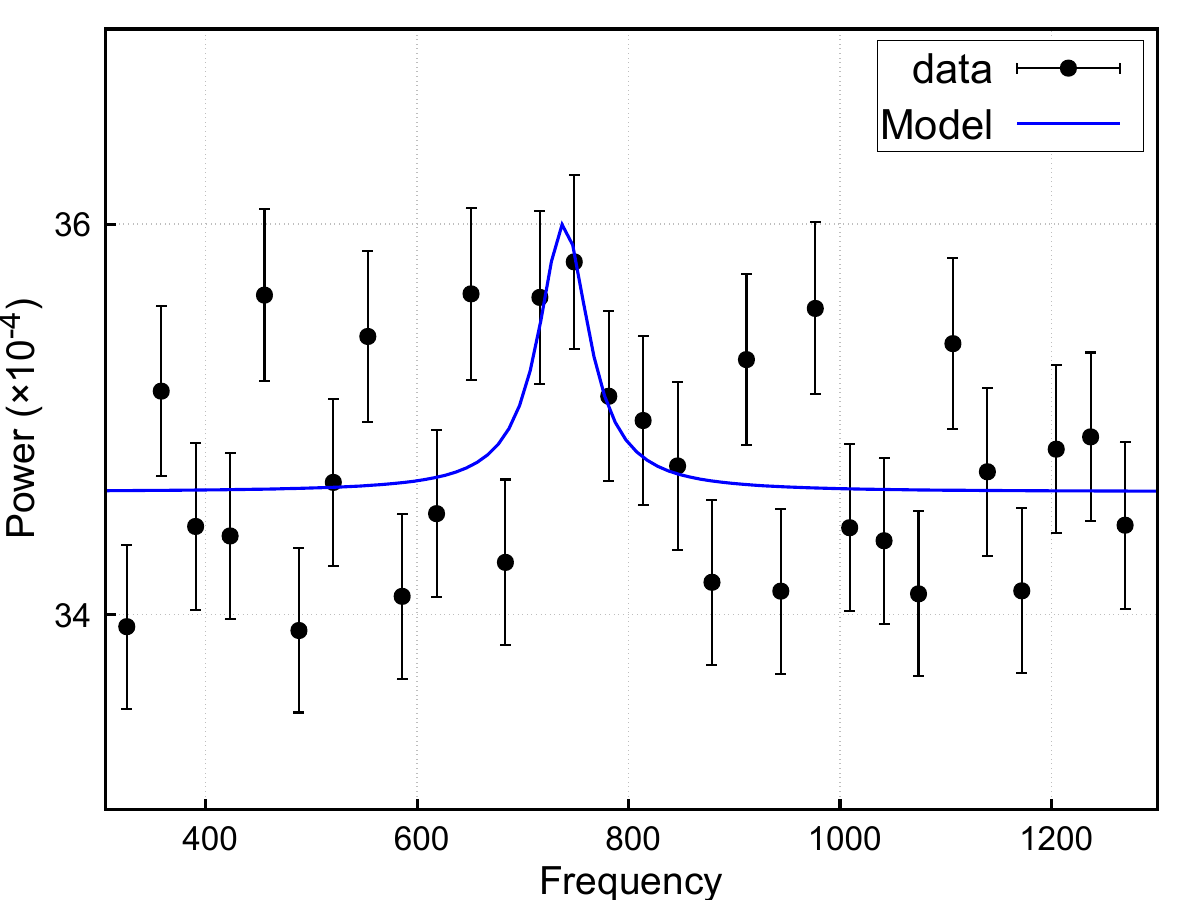} & \includegraphics[width=0.45\textwidth]{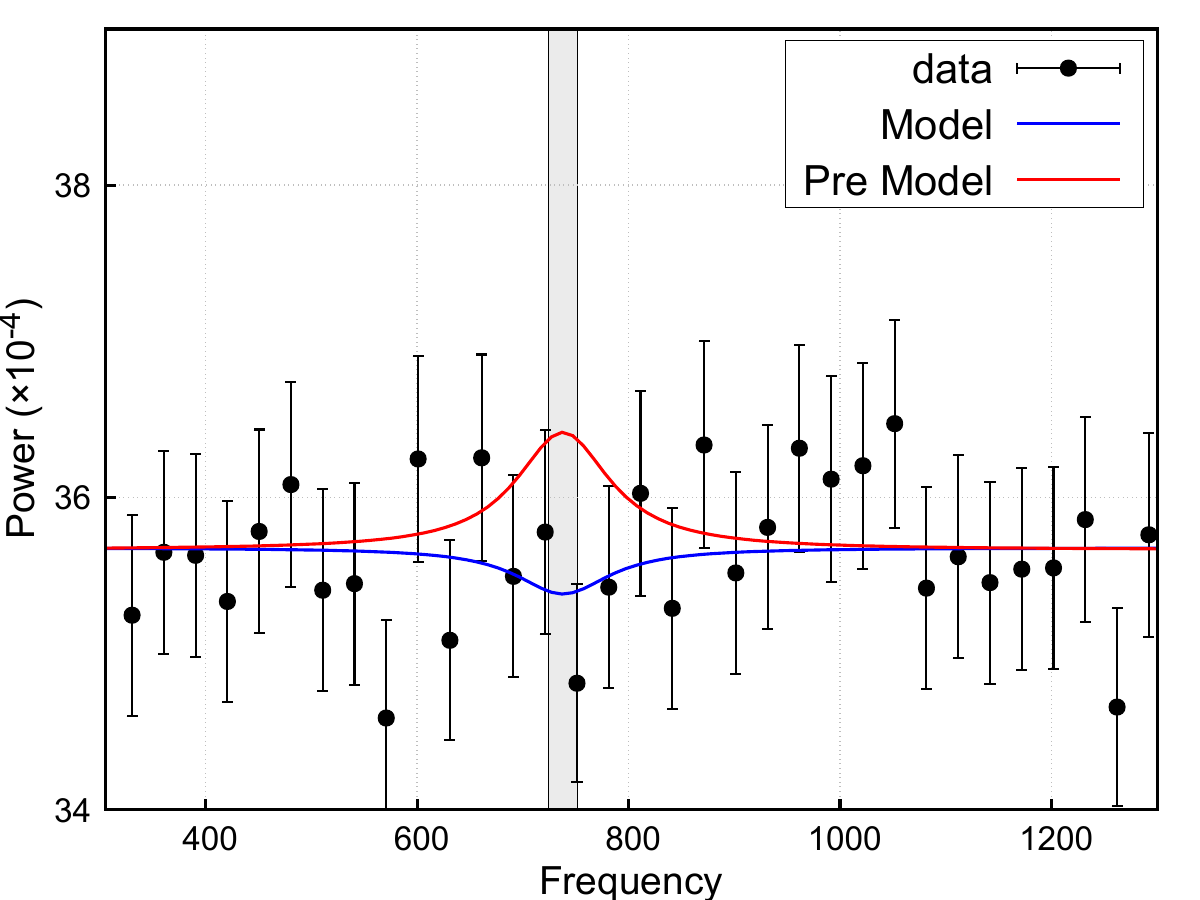}\\
Post 100\,--\,200\,sec & Post 200\,--\,300\,sec\\
\includegraphics[width=0.45\textwidth]{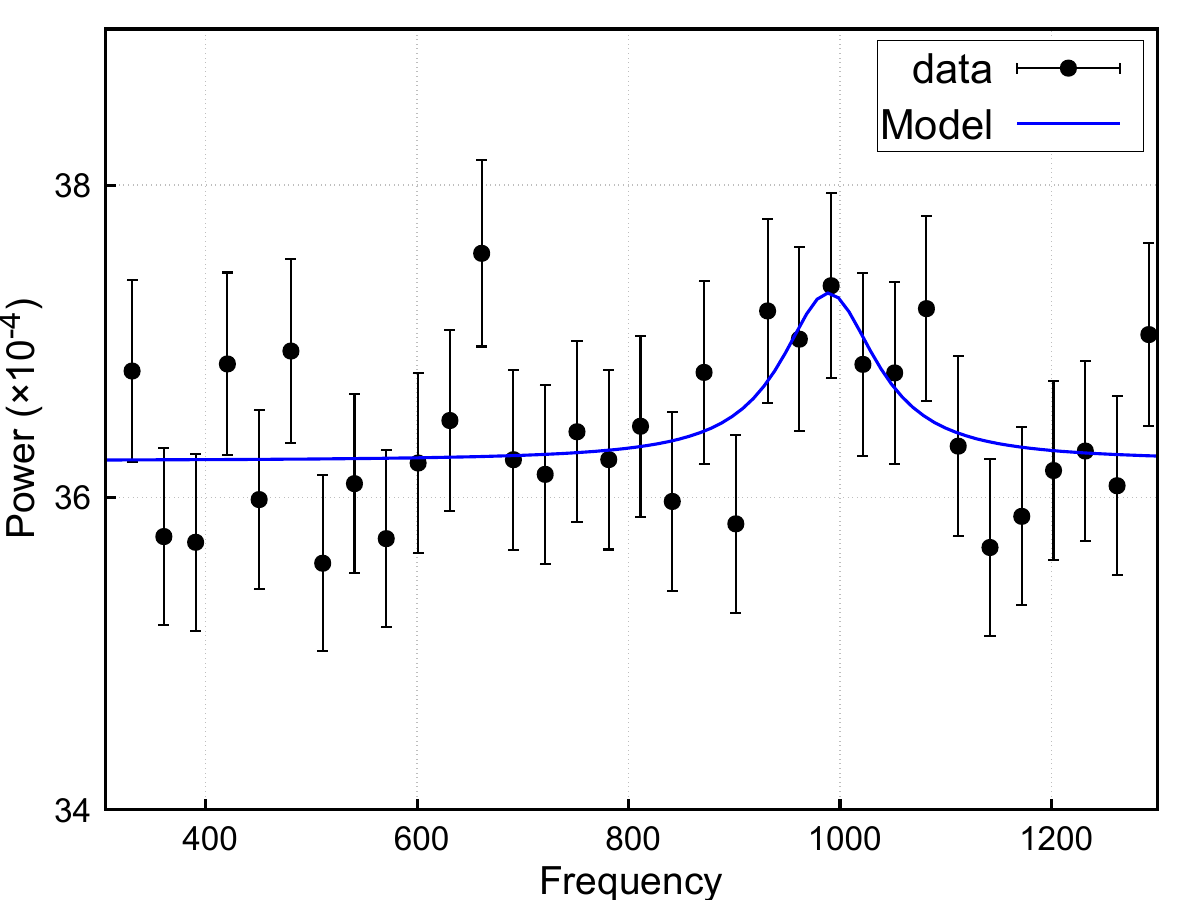} & \includegraphics[width=0.45\textwidth]{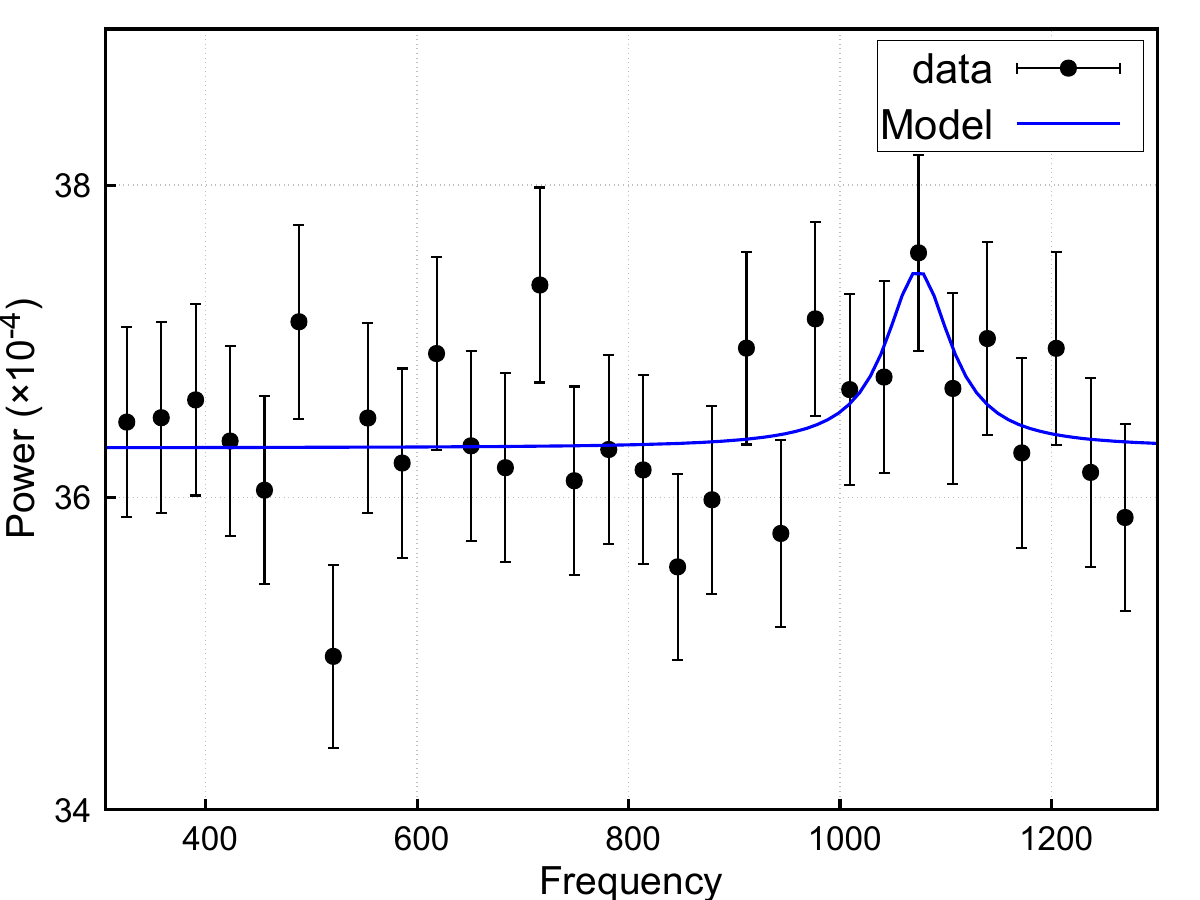}\\
Post 300\,--\,400\,sec & \\
\includegraphics[width=0.45\textwidth]{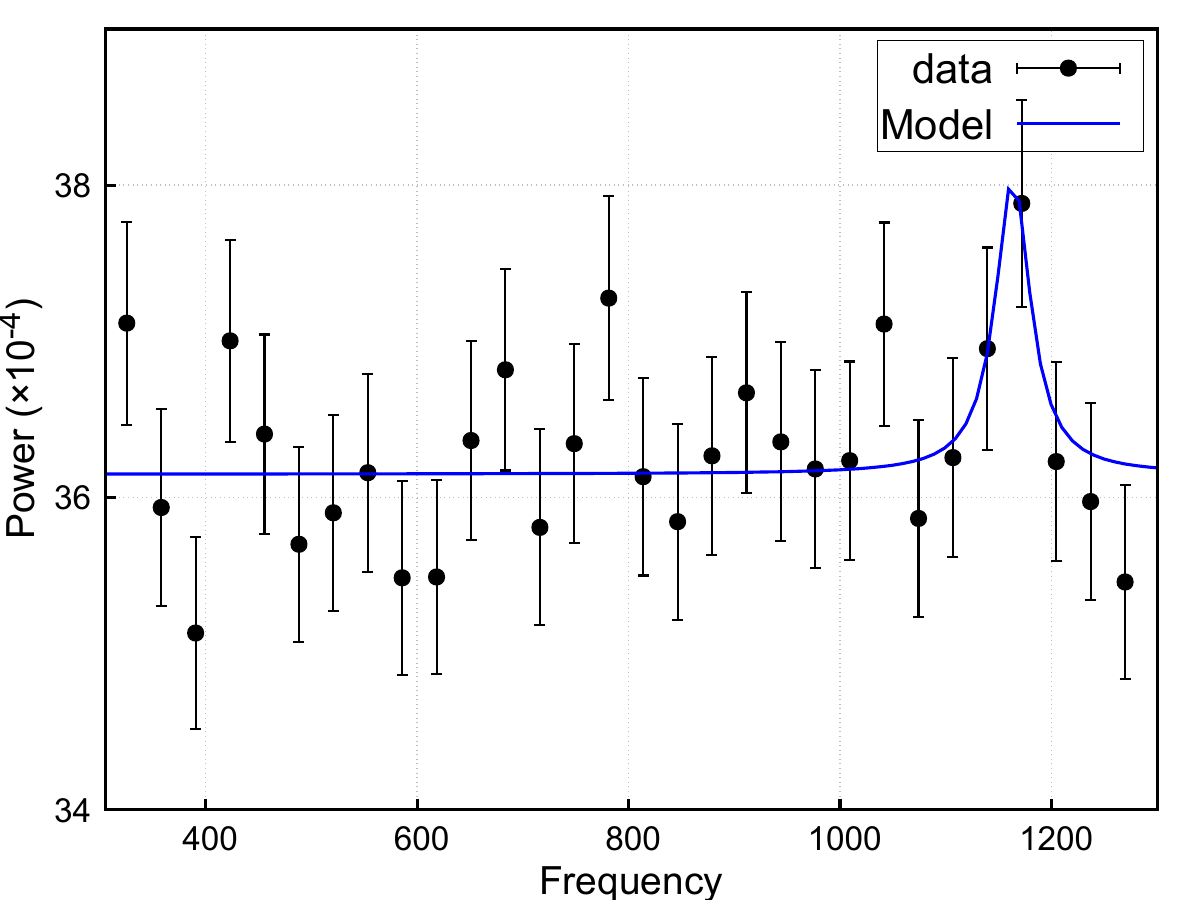} & \\
\end{tabular}
\caption{The evolution of the kHz QPO in the pre-burst and the post-burst zone of the Type--I TNB of Observation 2. While the kHz QPO is present in pre-burst data, the same QPO is not present in the post-burst within 0-100 sec data (we over-plot the pre-burst QPO with the blue line, highlighting the lack of signal). In some scenarios { (-100--0\,sec)}, we have changed the y-axis limit to show the prominent presence of the QPO, which has been marked with the *. The frequency is measured in Hz, and the power is rms normalized.}
\label{Obs2-evol}
\end{figure*}


\begin{figure*}
\begin{tabular}{c c}
Pre -200\,--\,0\,sec (*)& Post 0\,--\,200\,sec (*)\\
\includegraphics[width=0.45\textwidth]{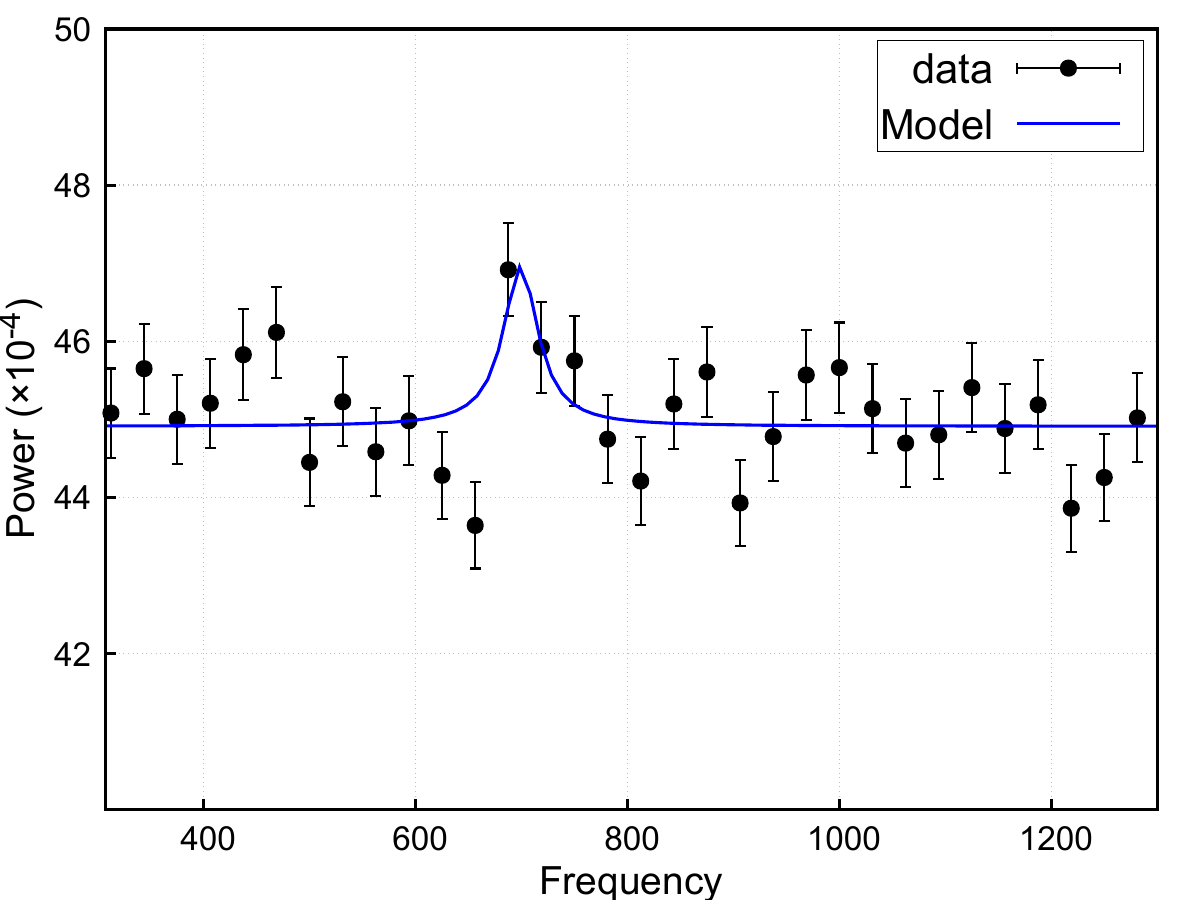} & \includegraphics[width=0.45\textwidth]{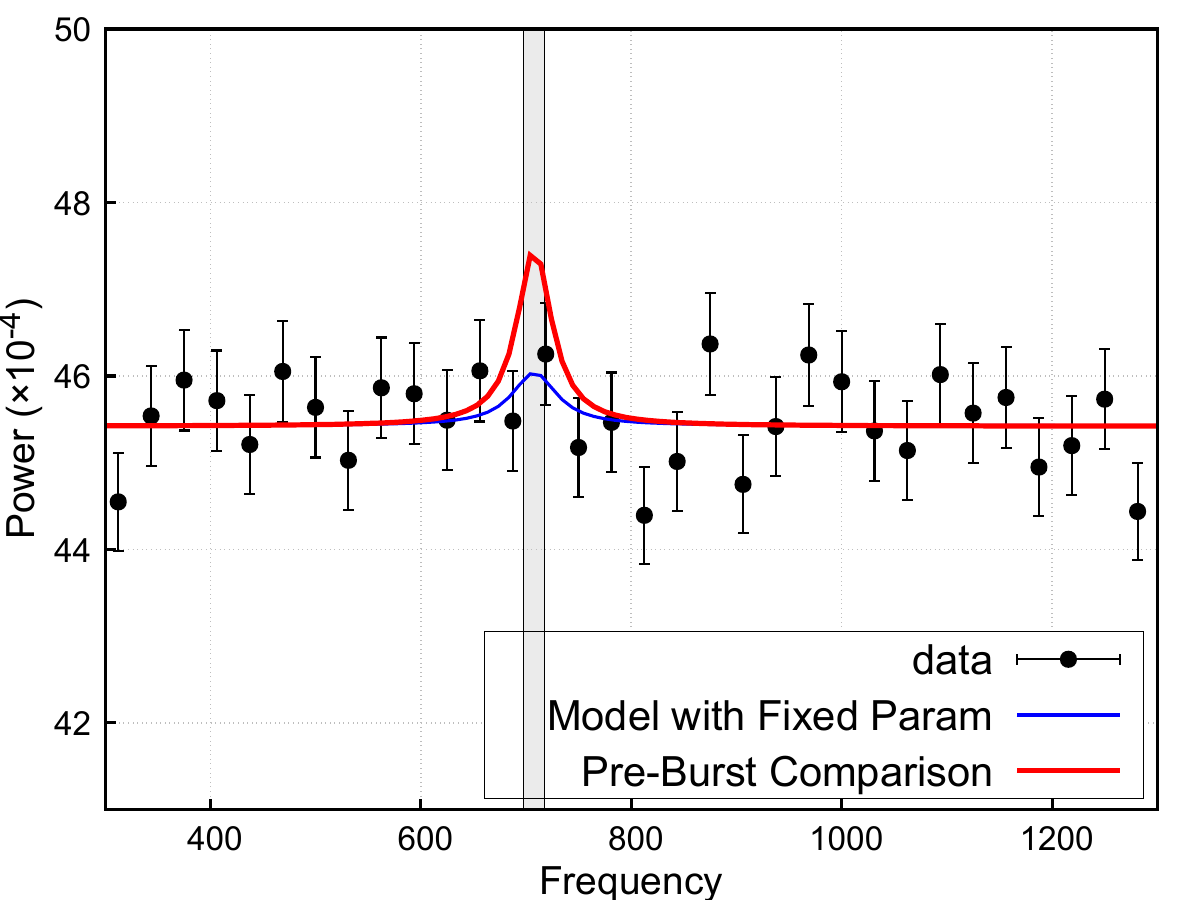}\\
Post 200\,--\,400\,sec & Post 400\,--\,600\,sec\\
\includegraphics[width=0.45\textwidth]{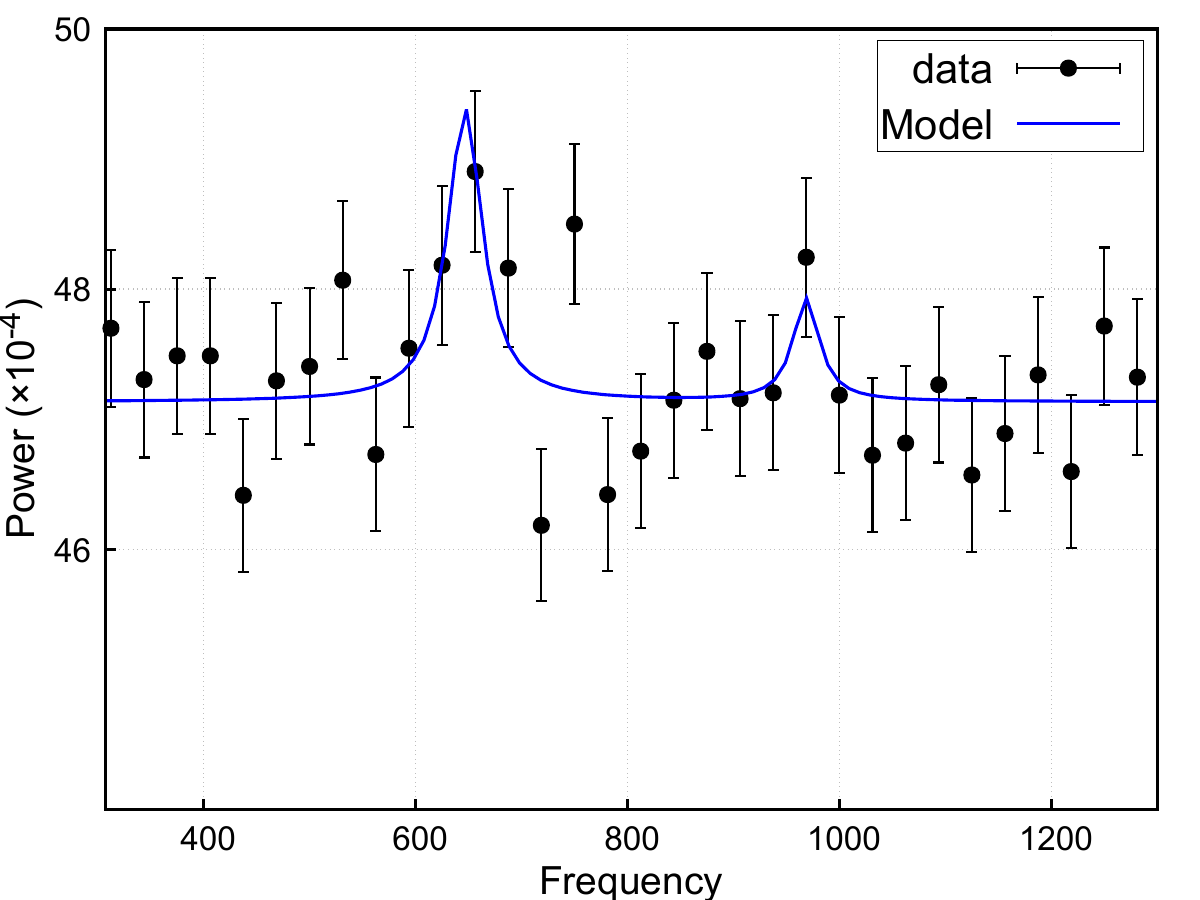} & \includegraphics[width=0.45\textwidth]{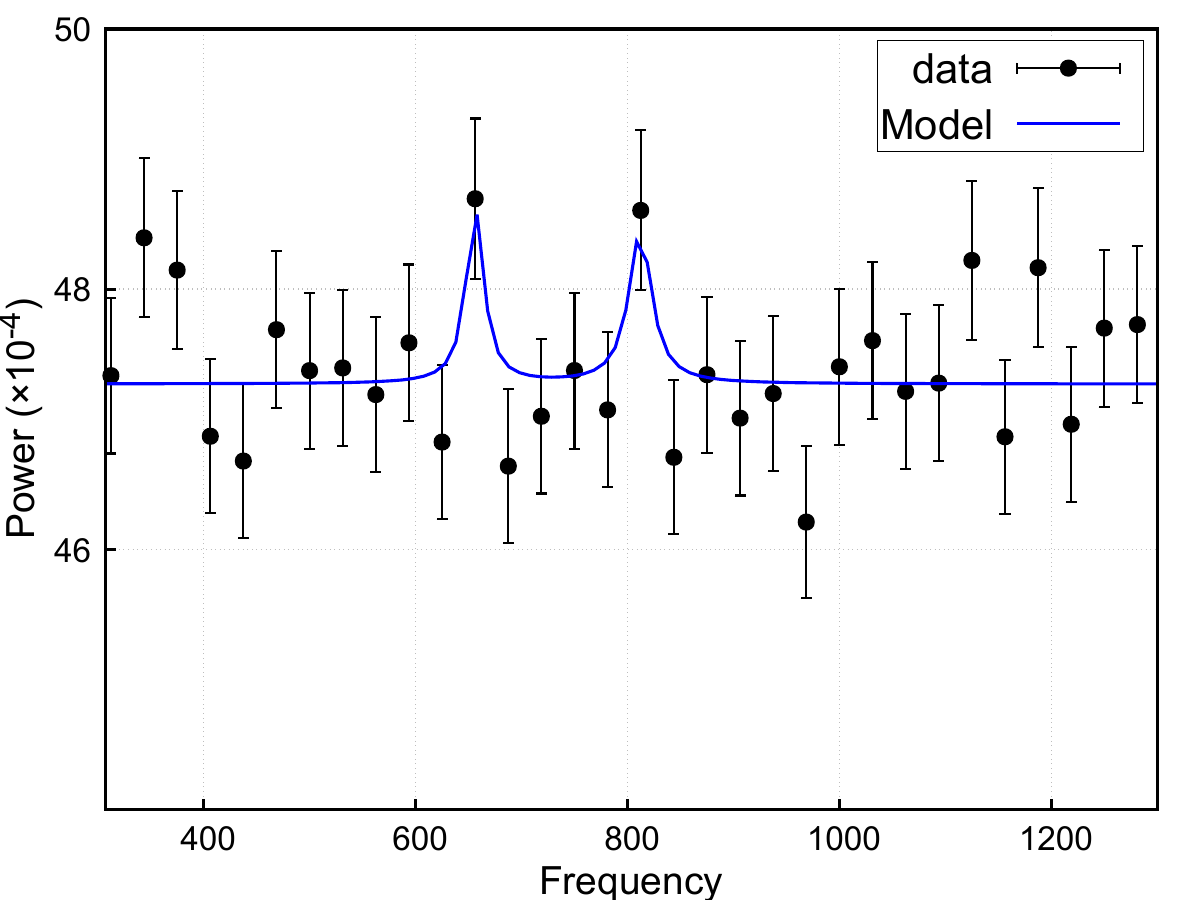}\\
Post 600\,--\,800\,sec & Post 800\,--\,1000\,sec\\
\includegraphics[width=0.45\textwidth]{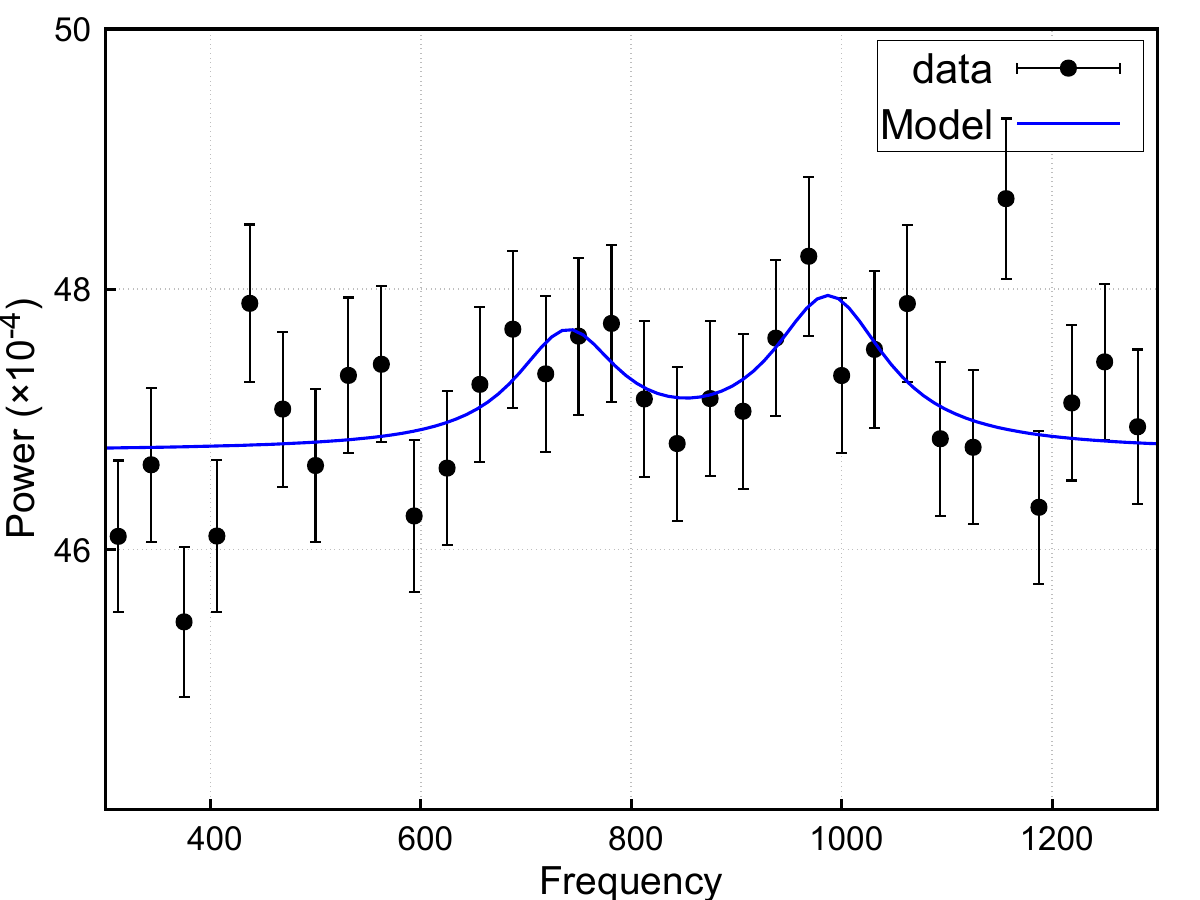} & \includegraphics[width=0.45\textwidth]{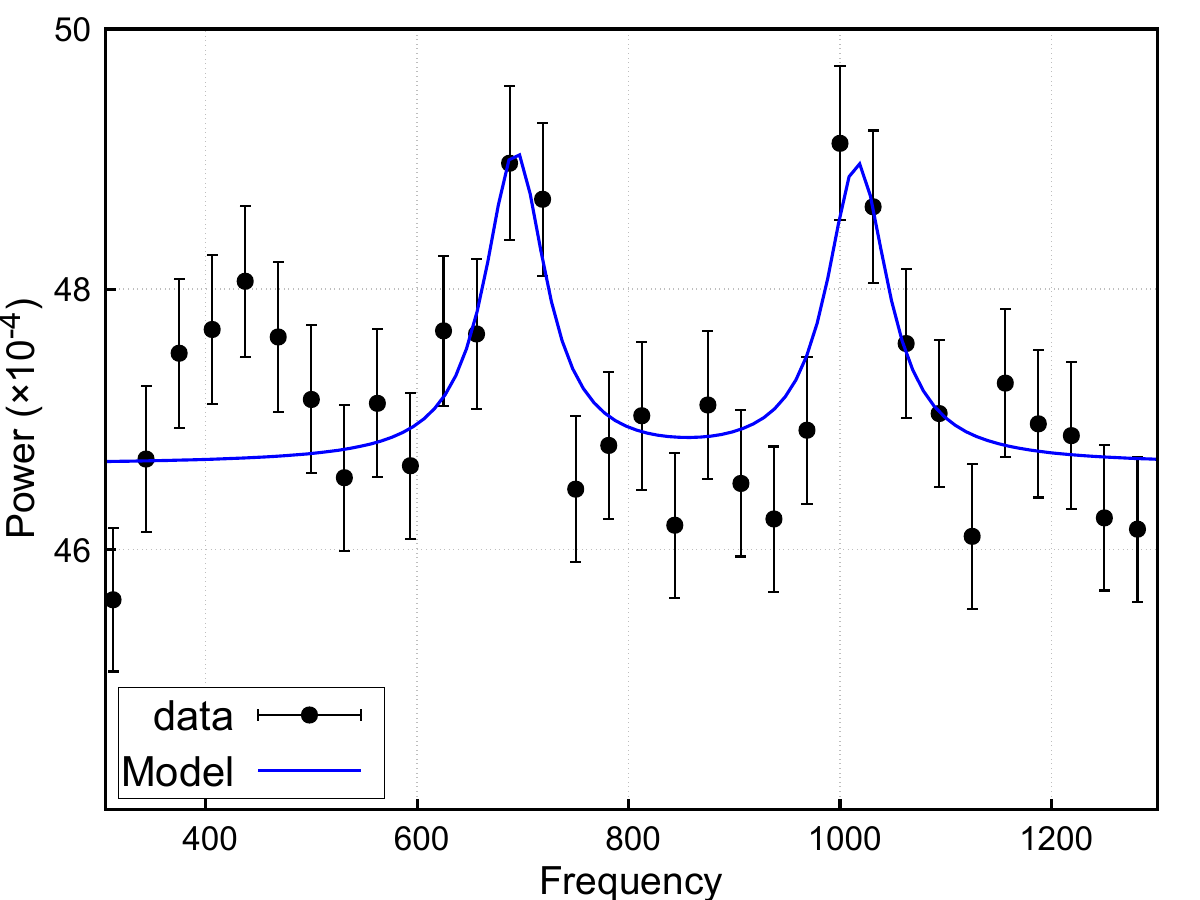}\\
\end{tabular}
\caption{The evolution of the kHz QPO in the pre-burst and the post-burst zone of the Type--I TNB of Observation 3. While the kHz QPO is present in pre-burst data, the same QPO is not present in the post-burst within 0-200 sec data (we over-plot the pre-burst QPO with the blue line, highlighting the lack of signal). In some scenarios {( -200--0\,sec, 0--200\,sec)}, we have changed the y-axis limit to show the prominent presence of the QPO, which has been marked with the *. The frequency is measured in Hz, and the power is rms normalized.}
\label{Obs3-evol}
\end{figure*}


In Figure~\ref{Fig00}, we show the source evolution through its hardness-intensity diagram (HID), constructed using the background-subtracted PCU 20 light curve with a bin size of 1024\,s. The hardness ratio is defined as the count rate ratio between the 8--20\,keV and 3--8\,keV energy bands, while the intensity represents the total counts in the full 3--20\,keV energy band. The source is primarily located in the softer region of the diagram, except during Observation 1. As expected, the source count rate increases during the softer state.

As highlighted in Figure \ref{Fig01}, the light curves of the three studied observations show the presence of Type--I TNBs across the three {\em AstroSat} energy bands. In Observation 1 and Observation 3, three Type--I TNBs are identified, while Observation 2 shows only a single Type--I TNB. To analyze the evolution of kHz QPOs before, during, and after the TNBs, a sufficient amount of data ($>$\,200\,s) is required to support any conclusions. Based on this criterion, two TNBs from Observation 3 and two TNBs from Observation 1 are excluded from further analysis due to insufficient data either preceding or following the TNB (see Figure \ref{Fig01}), with a single burst meeting our criteria in each of observation 1, 2, and 3, labeled B1, B2, B3, respectively, in Figure~\ref{Fig01}. Throughout this work, the periods before and after a TNB are referred to as the Pre-Burst Zone and Post-Burst Zone, respectively. For reference, in subsequent discussions from here on, we will call the B1 in Obs 1 as TNB--1, B2 in Obs 2 as TNB--2 and B3 in Obs 3 as TNB--3 (see figure \ref{Fig01}). The corresponding burst durations $\sim$\,120 seconds have been excluded from the light curves for further analysis. We further separately analyzed the Pre-Burst and Post-Burst Zones of each TNB. 


\section{Temporal Analysis \& Results} 
\label{subsec:temporal}

The LAXPC instrument onboard {\em AstroSat} provides an excellent opportunity to study the temporal behavior of the source \citep{Antia2017, klitzing:Antia_2021}. We utilized the standard routine {\tt laxpc\_find\_freqlag} to generate the PDS analyzed in this article \citep{Yadav2016}. This routine includes several configurable flags that allow users to tailor the PDS creation to their requirements. Among these, the {\tt -h} flag controls the maximum frequency of the PDS (Nyquist frequency), while the {\tt -l} flag specifies the minimum frequency or frequency resolution. For this study, we set the minimum frequency resolution to 31\,Hz and the Nyquist frequency to 2000\,Hz. All the PDS have been generated in the 3--20\,keV energy range (see Figure \ref{Obs1-evol}, \ref{Obs2-evol} and \ref{Obs3-evol}). To analyze the PDS, we have segregated the pre-Burst Zone and post-Burst Zone and {performed the search for the kHz QPO within 400--1200\,Hz as the maximum value of the upper kHz QPO reported for this source is $\sim$\,1176\,Hz \citep{1997ApJ...479L.141W}}. The pre-burst zone is defined as the 100--200\,sec interval before the burst onset, while the post-burst zone corresponds to the same duration after the burst, ensuring that the average count rates in both regions match (see Table \ref{Table3} and Figure \ref{Obs1-evol}, \ref{Obs2-evol} and \ref{Obs3-evol}). Further, the post-Burst Zone has been divided into several small segments to see the evolution of the kHz QPO after the Type--I TNB in this system. We have fitted all the PDS using either a single Lorentzian with a single constant or multiple Lorentzian with a constant. The Lorentzian are defined as

\begin{equation}
f(\nu) = \frac{A}{\pi} \left(\frac{\sigma}{(\nu-\nu_0)^2 + \sigma^2}\right)
\end{equation}
where A is the normalization of the Lorentzian, $\sigma$ is the width and $\nu_0$ is the centroid frequency. The best-fitted parameter values are tabulated in Table~\ref{Table3} for further reference. 

The results clearly show the presence of kHz QPOs in the Pre-Burst Zone across all scenarios. However, in the first 100--200\,s after the TNB, the PDS reveals no kHz QPOs. These QPOs reappear after approximately 200 \,s. As illustrated in Table \ref{Table3} and Figure \ref{Obs1-evol}, \ref{Obs2-evol} and \ref{Obs3-evol}, the kHz QPO frequency initially drops below the Pre-Burst frequency, then increases, approaching to the Pre-Burst value in 800--1000\,s after the burst. Additionally, the upper kHz QPO frequency differs significantly between the Pre-Burst and Post-Burst zones. In the B2 {zone}, there is no clear evidence of the lower kHz QPO within 200--500\,s Post-burst, and due to the lack of continuous data, it is unclear if it eventually surpasses the Pre-Burst frequency (see Figure \ref{Obs1-evol}, \ref{Obs2-evol} and \ref{Obs3-evol}).

The root mean square (RMS) amplitude of a kHz QPO  quantifies the strength of the oscillation and is proportional to the $\sqrt{\text{normalization}}$ of the Lorentzian \citep{belloni2002,vanderklis2006}. The RMS is often expressed as a percentage. We have computed the rms in the 3--20\,keV energy band, which is given in Table \ref{Table3}.
{To assess the significance of the detected frequencies, we conducted a statistical test by computing the signal-to-noise ratio (SNR) for each QPO using the method outlined in \citet{2004astro.ph.10551V,2008MNRAS.384.1519B}. Using this, we now obtain upper limits on the QPO RMS amplitudes when the SNR $<$ 3. Since this SNR does not account for the number of frequency bins that have been searched, we additionally estimated the null hypothesis probability for each QPO using the approach of \citet{2008MNRAS.384.1519B}, where we also perform a simulation study to confirm the results. Details have been added in the Appendix \ref{Appendix I}. The results indicate that for SNR $\geq$ 3, the null hypothesis probability when one considers number of frequency bins that are being searched, is still greater than 95\% confidence level.} Since the PDS in the post-Burst Zone during the initial 200\,s does not show a definitive QPO signature, we used the pre-Burst Zone parameters to compare the shift in the 3--20\,keV rms between the pre- and post-burst zones.

Table \ref{Table3} illustrates the evolution of the 3--20\,keV rms amplitudes for the lower and upper kHz QPOs in the pre- and post-burst zones corresponding to TNB--3 and TNB--1. In both cases, the lower kHz QPOs exhibit a notable drop of approximately 5--6\,\% in rms amplitude between the pre-burst phase and the initial 200\,s post-burst. Beyond this 200\,s mark, the rms increases and eventually approaches to the pre-burst level. For the upper kHz QPOs, no detection was made in the pre-burst zone, preventing direct comparison between the pre- and post-burst phases in TNB--3. However, for TNB--1, a similar pattern emerges: the rms decreases during the first 200\,s post-burst, followed by an increase of about 5--6\,\%. It is to be noted that for Observation 3, the third burst (OB4 see Figure \ref{Fig01}) has been omitted from the analysis. However, it meets the necessary conditions for the analysis, as we cannot detect any QPO in the pre-burst and post-burst zones.


\section{Discussion \& Conclusion}
\label{sec:discussion}
For over two decades, it has been hypothesized that Type--I thermonuclear bursts (TNBs) can influence, and in some cases entirely disrupt, the inner accretion flow in neutron star low-mass X-ray binaries (NSLMXBs) \citep[e.g.,][]{Zand2011, Keek2016, ZheYan2024}. Since the work of \citet{refId0}, there has been a lack of systematic observational studies addressing the full extent of this impact. In this study, we investigate the evolution of kHz QPOs before and after TNBs using X-ray observations in the 3--20\,keV band. Our results extend the findings of \citet{refId0}, revealing significant changes in QPO frequency and strength, strong indicators of inner disk disruption and subsequent reformation.

kHz QPOs provide a sensitive probe of the inner accretion flow. Several models have been proposed to explain their origin, including the relativistic precession model \citep[RPM;][]{1998ApJ...492L..59S,1999PhRvL..82...17S}, tidal disruption model \citep{vcadevz2008tidal,2009AIPC.1126..367G}, and disk oscillation models \citep{kato2007frequency}. Although no consensus has been reached, \citet{2011ApJ...726...74L} showed that the tidal disruption model provides a better fit than the RPM for 4U,1636$-$536, with lower \(\chi^2/\text{dof}\) and improved parameter constraints. However, it fails to explain the expected drop in lower kHz QPOs at high frequencies. In contrast, recent work by \citet{chattopadhyay2025spectraltiminganalysiskilohertzquasiperiodic} suggests that the RPM better fits the same dataset, yielding a mass of $\sim$2.37 M$_\odot$ and an angular momentum of $\sim$1.40\,I$_{45}$/M$_\odot$. Additional studies considering radiative mechanisms \citep{lee2001compton,kumar2014energy,kumar2016constraining,2022MNRAS.515.2099B,Chattopadhyay_2024} place the origin of kHz QPOs within 12--15\,km of the neutron star, linking them closely to the innermost regions of the accretion disk.

Since TNBs originate from the neutron star surface \citep{1993SSRv...62..223L,2003astro.ph..1544S,Galloway_2008,Galloway2021}, the intense radiation during a burst is expected to irradiate and potentially disturb the nearby inner accretion flow \citep[e.g.,][]{Zand2011, Keek2016, ZheYan2024}. This interaction can lead to observable changes in kHz QPO properties. \citet{Yu_1999} and \citet{refId0} provided early evidence of such interactions, including cases where QPOs disappear following intense bursts.

\citet{refId0} reported two types of TNB-related QPO behavior in 4U\,1636$-$536. In long bursts, with peak count rates reaching $\sim10^{4}$\,counts\,$s^{-1}$ (6--50\,keV), QPOs vanished and reappeared after delays of up to several hundred seconds. In other bursts, QPOs were largely unaffected. Their analysis, limited to -200 to +400\,s around the burst, found no QPOs for 80--248\,s post-burst in some cases.

Our study (see Section \ref{subsec:temporal}) expands the temporal coverage beyond 400\,s and detects persistent lower kHz QPOs in all pre-burst phases (TNB--1, TNB--2, TNB--3), with frequencies ranging from 683--768\,Hz. The upper kHz QPO is not detected in pre-burst. In post-burst, both QPOs vanish for the first 100--200\,s--consistent with the upper limits set by \citet{refId0}--indicating either a disruption of the inner disk or the dominance of burst emission suppressing QPO visibility. This is supported by:

\begin{itemize}
    \item Non-detection of QPOs in 10--20\,keV, where burst intensity is lower, ruling out energy band dependence.

    \item Uniform $\sim$\,100--200\,s QPO reappearance delay across all three TNBs, regardless of burst peak intensity or spectral state.
\end{itemize}

After $\sim$200\,s, kHz QPOs reappear. In TNB--3, the lower kHz QPO fractional rms gradually increases between 200--800\,s, exceeding the pre-burst amplitude by 800--1000\,s. A similar trend is seen in TNB--2, although the lower QPO is undetected between 200--600\,s. Interestingly, the upper kHz QPO re-emerges $\sim$200\,s post-burst even if it was absent before, hinting at a potentially earlier recovery of high-frequency modes. 

{To explain this one may interpret burst-induced disruptions as temporarily removing an inner flow and as the disk recovers post-burst, the inner accretion flow re-forms, possibly explaining the gradual reappearance of higher-frequency QPOs. The inner regions of the accretion flow is generally associated with the kHz QPO as for example in the model where they are attributed to the Keplerian and the Epicyclic frequency of a characteristic radius close to the neutron star in the RPM model \citep{1998ApJ...492L..59S,1999PhRvL..82...17S}. The low frequency QPO could then arise from the  Lense–Thirring precession of the inner flow \citep{10.1111/j.1745-3933.2009.00693.x,10.1111/j.1365-2966.2010.16614.x}. The disappearance and re-appearance of the kHz QPOs, would be expected from models where they arise from an inner accretion flow which gets disrupted during a TNB.
}

The reappearance timescale of QPOs likely reflects the viscous timescale ($t_{\text{visc}}$) governing disk refilling after burst-induced disruption \citep{1995ASIC..450..355S}. Viscous timescale is estimated as $t_{\text{visc}} \sim \frac{R_{\text{in}}^2}{\nu}$., where R is the radial distance and $\nu$ the kinematic viscosity \citep{klitzing:frank1985accretion,ingram2012physical}. During a burst, the radiation pressure at near-Eddington luminosities ($L_{Edd}\approx3\times10^{38}$\,erg/s for a 2.4 M$_\odot$ NS) can exceed gravity, pushing the disk outward \citep{fragile2020interactions}. Afterwards, the inner disk gradually refills over $t_{\text{visc}}$, depending on $\nu$.
Assuming an inner disk radius of \( R_{\text{in}} = 4 \times 10^6 \, \text{cm} \) \citep{chattopadhyay2025spectraltiminganalysiskilohertzquasiperiodic} and a kinematic viscosity in the range of \( \nu \sim 10^{10} - 10^{13} \, \text{cm}^2/\text{s} \) \citep{klitzing:frank1985accretion}, we estimate:

For \( \nu = 10^{10} \, \text{cm}^2/\text{s} \):

\begin{equation}
t_{\text{visc}} = \frac{(4 \times 10^6)^2}{10^{10}} = \frac{1.6 \times 10^{13}}{10^{10}} = 1600 \, \text{s} .
\end{equation}

For \( \nu = 10^{11} \, \text{cm}^2/\text{s} \):

\begin{equation}
t_{\text{visc}} = \frac{(4 \times 10^6)^2}{10^{11}} = \frac{1.6 \times 10^{13}}{10^{11}} = 160 \, \text{s} .
\end{equation}

For \( \nu = 10^{13} \, \text{cm}^2/\text{s} \):

\begin{equation}
t_{\text{visc}} = \frac{(4 \times 10^6)^2}{10^{13}} = \frac{1.6 \times 10^{13}}{10^{13}} = 1.6 \, \text{s}.
\end{equation}

\vspace{5mm}

The 160--200 s delay we observe corresponds well with $\nu\sim10^{11}$\,cm$^{2}$/s, supporting the interpretation that viscous timescales govern QPO recovery after burst-induced disk disruption.

In summary, our results provide compelling evidence for temporary inner disk disruption caused by Type--I TNBs, followed by viscous refilling. The systematic disappearance and reappearance of kHz QPOs—along with changes in their strength—support a strong causal link between bursts and inner accretion flow dynamics. 

\subsection{Caveats and Limitations}

While our study presents compelling evidence for the disruption and reformation of the inner accretion disk by Type--I thermonuclear bursts (TNBs) in 4U\,1636--536, several caveats must be considered when interpreting the results:

First, our analysis is based on a small sample of bursts (TNB–1, TNB–2, and TNB–3). Although these provide valuable insights, the limited number restricts the statistical robustness and generalizability of our conclusions. Previous studies have shown that burst–QPO interactions can vary significantly from burst to burst \citep[e.g.,][]{Yu_1999, refId0}, highlighting the need for a broader sample to assess systematic trends.

Second, the non-detection of kHz QPOs in the immediate post-burst phase may result not only from physical disruption but also from instrumental limitations. High photon flux during bursts can increase background noise and mask subtle oscillatory features, especially in the power density spectra \citep[PDS;][]{Zand2011, Keek2016}. Therefore, the absence of QPOs might partially reflect observational constraints rather than true disappearance.

Third, the viscous timescale estimates we adopt are based on simplified $\alpha$--disk models \citep{klitzing:frank1985accretion, 1995ASIC..450..355S, ingram2012physical}. In reality, the accretion disk viscosity, structure, and response to bursts may be influenced by additional effects such as magnetic fields, burst-driven turbulence, or changes in disk ionization \citep{fragile2020interactions}. Hence, while our estimated recovery time aligns with the observed QPO reappearance ($\sim$\,200\,s), this should be regarded as an order-of-magnitude consistency rather than a precise measurement.

Fourth, the theoretical interpretation of kHz QPOs remains model-dependent. While our findings support the relativistic precession model \citep{1998ApJ...492L..59S, 1999PhRvL..82...17S}, alternative models such as the tidal disruption \citep{2011ApJ...726...74L} or disk oscillation frameworks \citep{kato2007frequency} may also reproduce certain aspects of the observed QPO behavior. As no single model currently explains all QPO characteristics across different sources, caution is warranted in drawing firm physical conclusions based solely on QPO properties.

Lastly, our study does not consider the role of the X-ray corona or jet in modulating or responding to burst events \citep{Russell2024}. These components are dynamically coupled with the disk and could potentially affect the QPO detectability and energy-dependent characteristics \citep{Stiele2013, 2022MNRAS.515.2099B}. Future multi-wavelength and higher-resolution studies may help disentangle these contributions.

\appendix
\section{Estimation of null hypothesis probability for a kHz QPO detection}
\label{Appendix I}
{We provide a formalism to estimate the null hypothesis probability for a kHz QPO detection by taking into account the number of frequency bins (i.e. the range of frequencies) which have been searched. This formalism is based on the results shown in \citet{1983ApJ...266..160L} and \citet{2004MNRAS.354..945M}.}

{Consider a system characterized by white (or Poisson) noise, such that the power density spectrum is frequency independent and given by \( P_0 \). For a finite number of segments, $M$, the sum of the power obtained for each segment will be distributed as a $\chi^2$ distribution with degree \(2M\) \citep{1983ApJ...266..160L}. This means that the probability distribution for the average computed power, \(P\), at a given frequency bin will be,}

{
\begin{equation}
\mathrm p\,(P)\,dP = \frac{2M}{P_0} \, \chi^2\left(\frac{2MP}{P_0}, 2M\right)\,dP
\end{equation}
Note that \( p\,(P)\,dP \) is properly normalized (i.e., its integral over all \( P \) is unity), and that the expectation value is given by}

{
\begin{equation}
<P> = \int P \times {\it p}(P) dP = P_0
\end{equation}
The probability that the computed power \( P \) is less than a given value \( P_{\text{max}} \) is}

{
\begin{equation}
\mathrm{p}\,(< P_{\text{max}}) = \displaystyle \int_0^{\frac{2M P_{\text{max}}}{P_0}} \chi^2(x, 2M)\, dx = \Gamma_I\left(M, \frac{M P_{\text{max}}}{P_0}\right)
\end{equation}}
{where \( \Gamma_I(a, x) \) is the incomplete gamma function. The probability for \( N \) frequency bins  that all the powers are less than \( P_{\text{max}} \) is then \( \mathrm p\,(< P_{\text{max}})^N \). This then provides the null hypothesis probability, that in the absence of a signal, the maximum value of the power for \( N \) number of frequency bins, \( P_{\text{max}} \), has been obtained by chance is}

{
\begin{equation}
\mathrm p_N\,(< P_{\text{max}}) = 1 - \left[\Gamma_I\left(M, \frac{M P_{\text{max}}}{P_0}\right)\right]^N
\end{equation}
}
{This probability will depend on the range of frequencies which are searched for and also on the frequency width. If there is a real detectable QPO in the light curve, then the probability will decrease to a significant minimum value when the frequency width is roughly equal to the width of the QPO. Thus, the formalism would be to compute ${\mathrm p_N}(<P_{max})$ for different frequency widths and if it decreases below a significant level, then the technique would have detected a QPO at the frequency at which $P_{max}$ is achieved with a width roughly corresponding to the frequency width.}

\begin{figure}
	\centering
	\begin{tabular}{c c}
    Panel 1 & Panel 2 \\
        \includegraphics[scale=0.36]{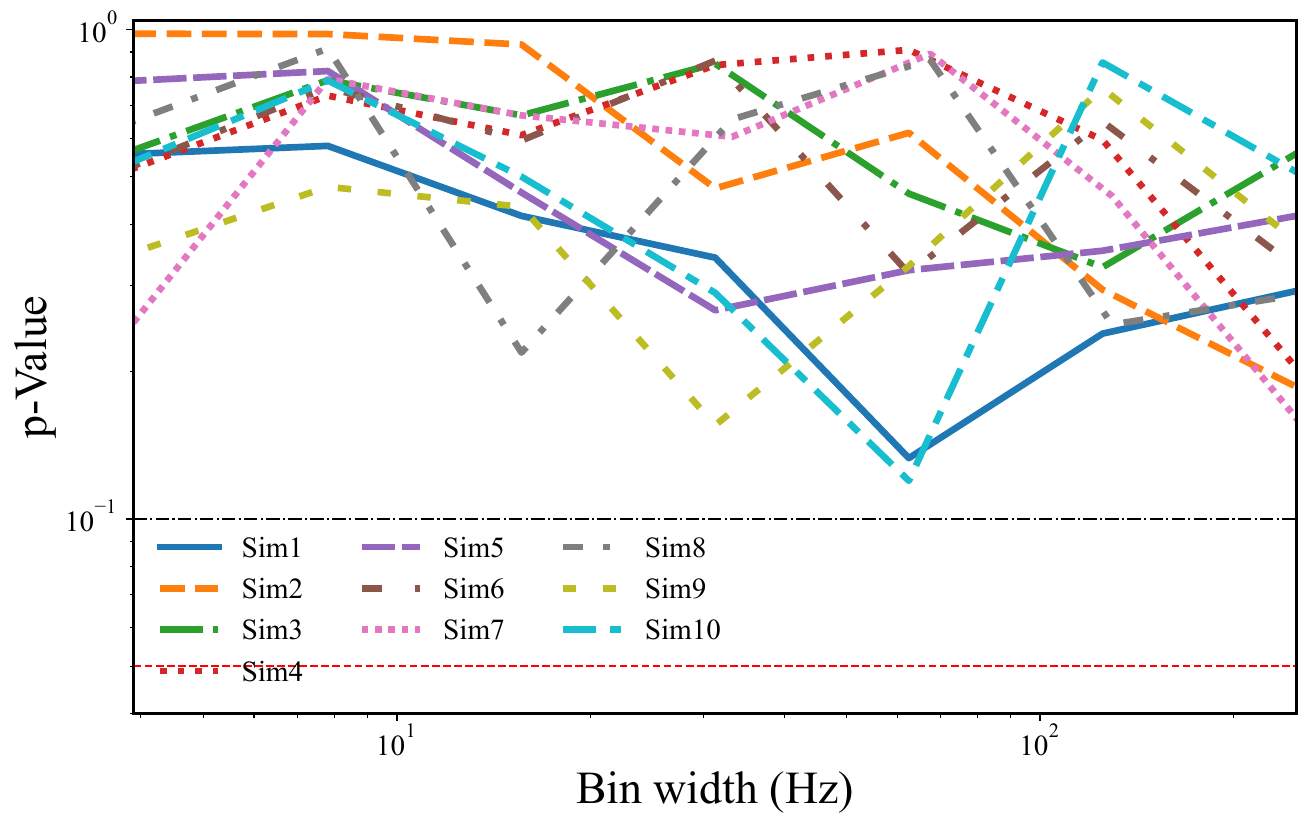} &
        \includegraphics[scale=0.36]{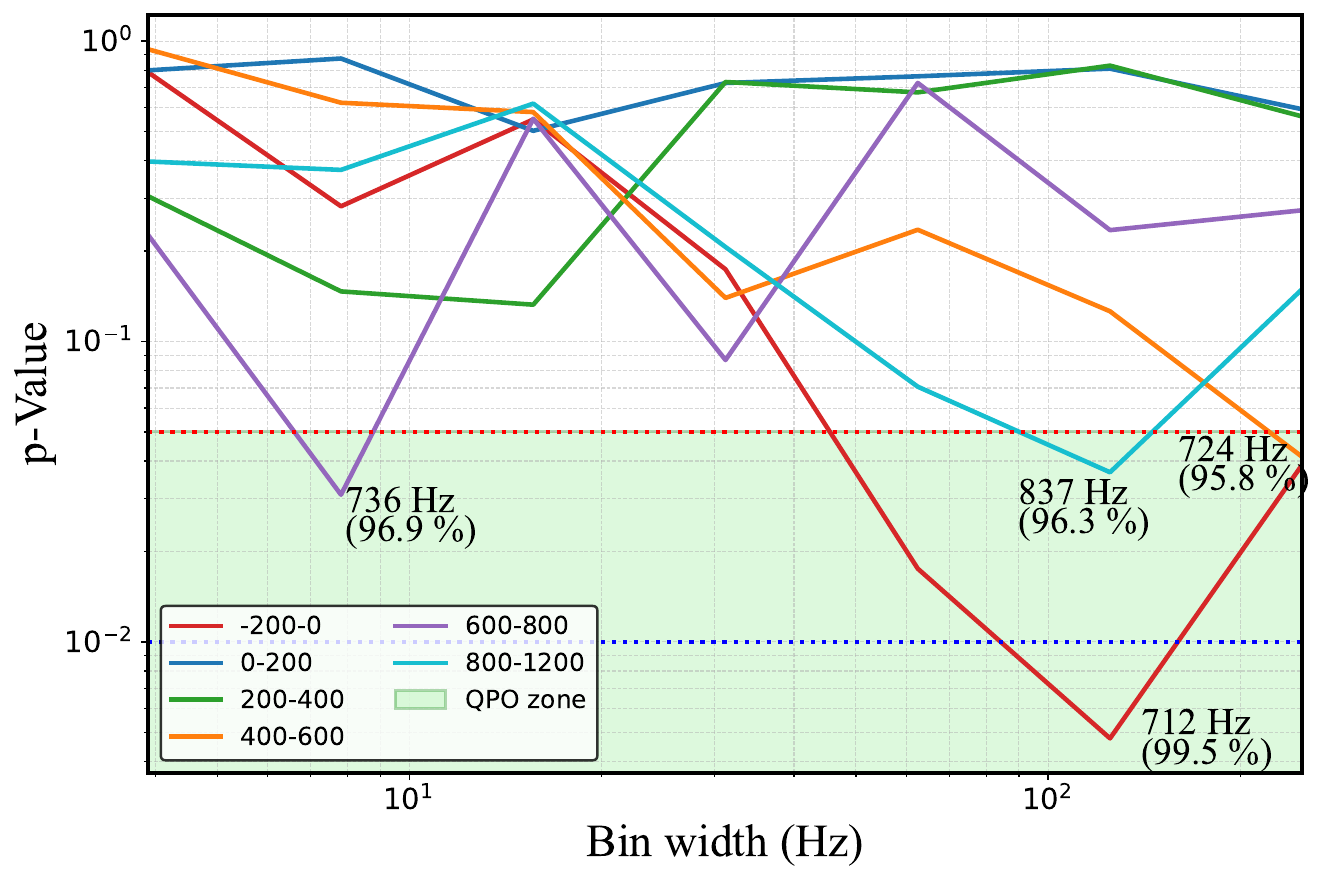}\\
    Panel 3 & Panel 4 \\    
        \includegraphics[scale=0.36]{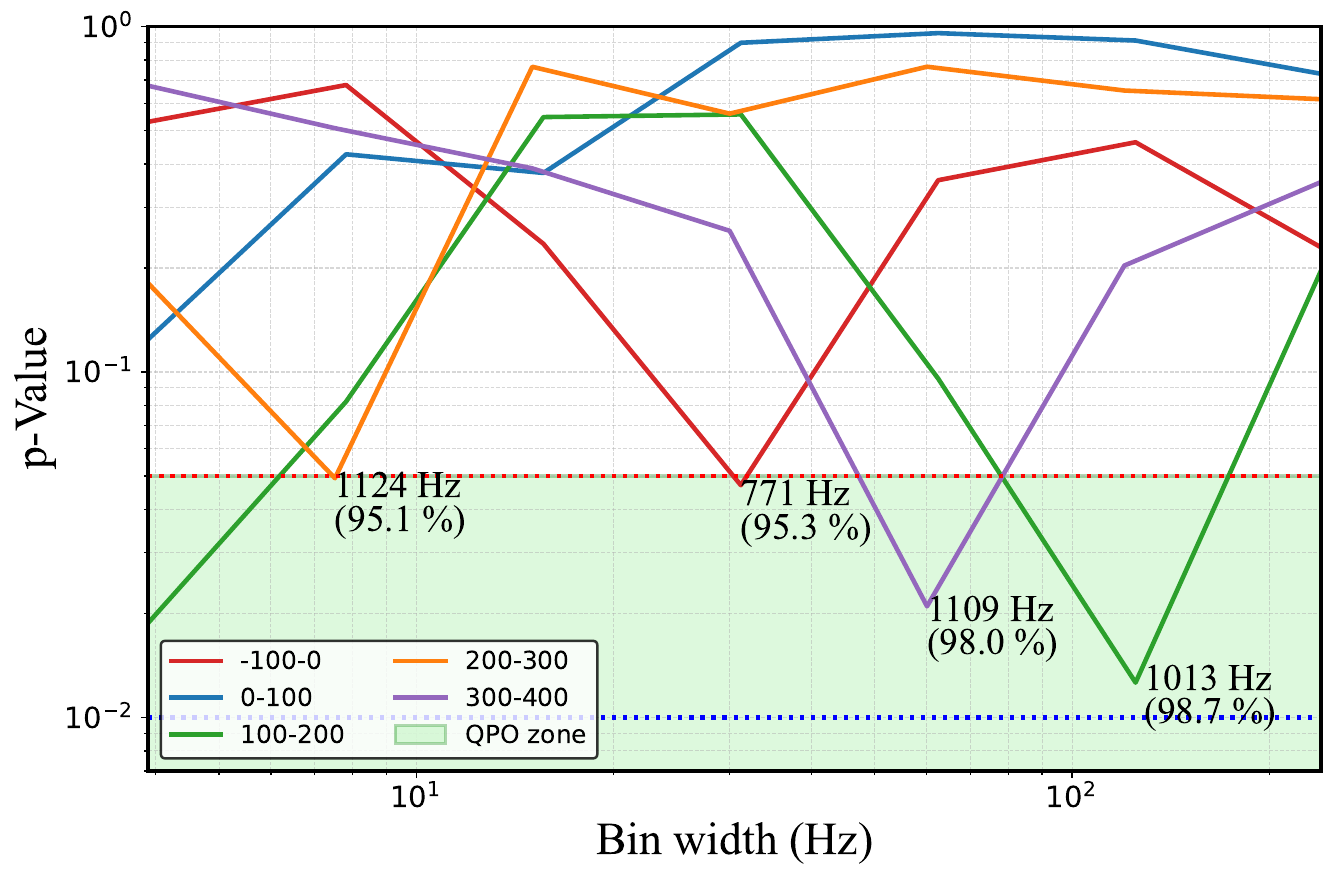} &
        \includegraphics[scale=0.36]{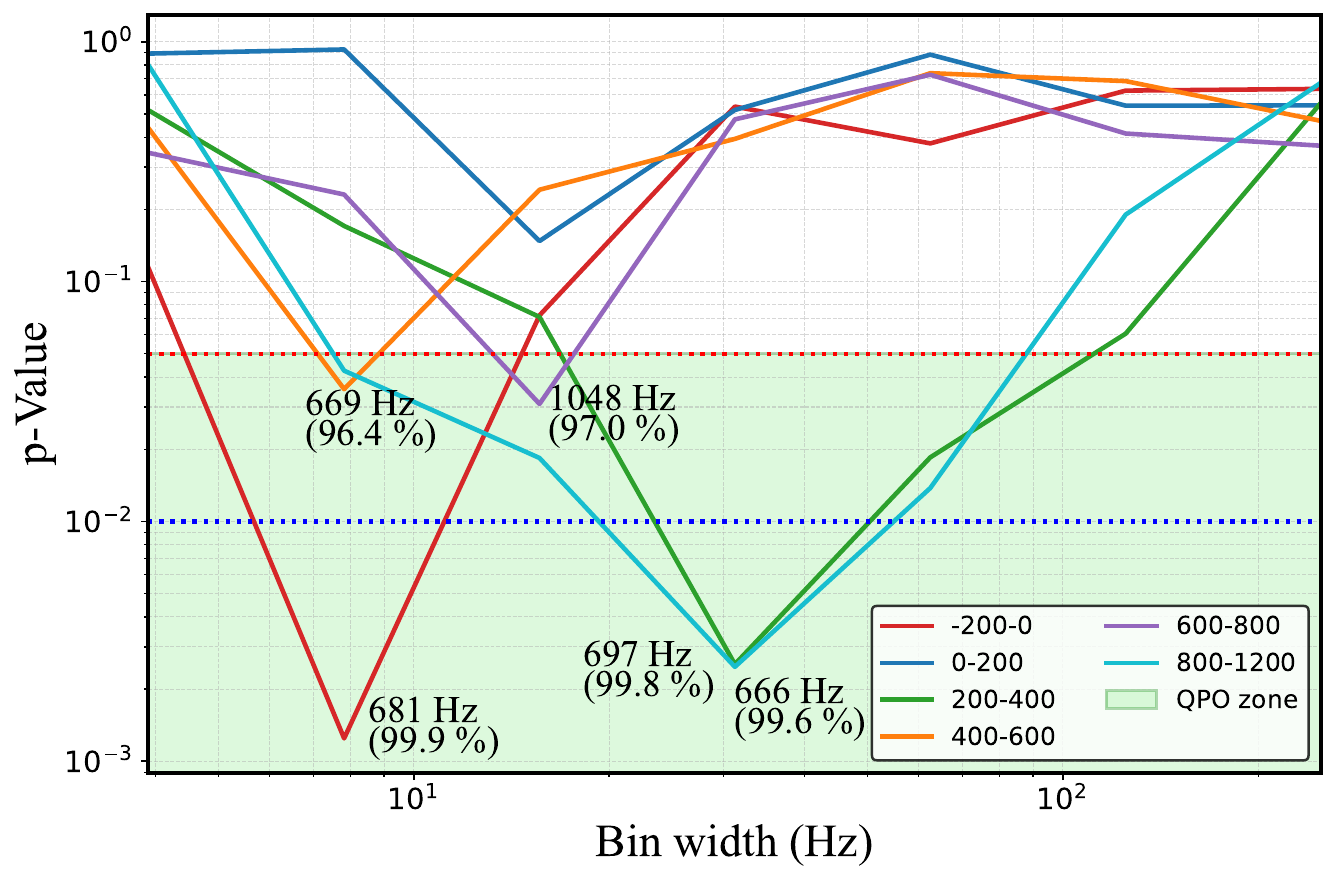} \\
    \end{tabular}
\caption{\footnotesize Plot of probability (p-value) vs. bin frequency. Black, red and blue horizontal lines mark the $p$-value thresholds of 0.1, 0.05 and 0.01. Each curve from Panel 2 to Panel 4 (left to right) corresponds to a segment from Obs 1 (Panel 2), Obs 2 (Panel 3) and Obs 3 (Panel 4) around Burst 1, 2, and 3, respectively. Panel 1  shows the simulated PDS in absence of intrinsic variability which are created using simulated event files that takes same spectra and count rate as the data . In such cases, ${\mathrm p_N} (< P_{max})$ is always less than 0.1. }    
\label{App0}
\end{figure}    


\begin{table*}[t]
\centering

\setlength{\tabcolsep}{12pt}
\renewcommand{\arraystretch}{1.15}
\begin{tabular}{c c c c c c c c c }
\hline
Freq & Bin Width & $p$ & $(1{-}p)\%$ & SNR & Fit Freq & Width & Obs & Segment \\
(Hz) & (Hz) &  &  &  & (Hz) & (Hz) &  & \\
\hline
712 & 125 & 0.0047 & 99.5 & 6.7 & 708 $\pm$ 13 & 42 $\pm$ 20 & 1 & -200--0 \\
626 & 7.81 & 0.51 & 49 & 1.1 & $708^{\dagger}$ & $42^{\dagger}$ & 1 & 0--200 \\
751 & 15.62 & 0.13 & 87 & 2.6 & $708^{\dagger}$ & $42^{\dagger}$ & 1 & 200--400 \\
724 & 250 & 0.0412 & 95.8 & 3.6 & 719 $\pm$ 15 & $<$61 & 1 & 400--600 \\
736 & 7.81 & 0.0308 & 96.9 & 4.45 & 785 $\pm$ 33 & $<$25 & 1 & 600--800 \\
\hline
619 & 62.5 & 0.0706 & 93 & 4.0 & 623 $\pm$ 13 & 30 $\pm$ 14 & 1 & 800--1200 \\
837 & 125 & 0.0366 & 96.3 & 4.3 & 811 $\pm$ 13 & 26 $\pm$ 12 & 1 & 800--1200 \\
\hline
\hline
771 & 31.25 & 0.047 & 95.3 & 3.3 & 738 $\pm$ 14 & $< 54 $ & 2 & -100--0 \\
642 & 15.62 & 0.37 & 63 & 0.5 & 738$^{\dagger}$ & 54$^{\dagger}$ & 2 & 0--100 \\
1013 & 125 & 0.013 & 98.7 & 4.41 & 989 $\pm$ 21 & 55 $\pm$ 35 & 2 & 100--200 \\
1124 & 7.81 & 0.049 & 95.1 & 3.1 & 1073 $\pm$ 20 & $37 \pm 26$ & 2 & 200--300 \\
1109 & 62.5 & 0.02 & 98 & 4.3 & 1163 $\pm$ 10 & $<$30 & 2 & 300--400 \\
\hline
\hline
681 & 7.81 & 0.0012 & 99.9 & 7.5 & 698 $\pm$ 6 & $<$13 & 3 & -200--0 \\
666 & 7.81 & 0.14 & 86 & 1.54 & $698^{\dagger}$ & $13^{\dagger}$ & 3 & 0--200 \\
666 & 31.25 & 0.0037 & 99.6 & 6.58 & 648 $\pm$ 10 & $<$15 & 3 & 200--400 \\
669 & 7.81 & 0.0356 & 96.4 & 3.0 & 655 $\pm$ 55 & $<$19 & 3 & 400--600 \\
1048 & 15.62 & 0.03 & 97.0 & 3.8 & 988 $\pm$ 27 & $<$66 & 3 & 600--800 \\
\hline
697 & 31.25 & 0.0024 & 99.8 & 7.13 & 695 $\pm$ 30 & $<$35 & 3 & 800--1200 \\
1001 & 15.62 & 0.018 & 98.2 & 5.3 & 1015 $\pm$ 8 & $<$35 & 3 & 800--1200 \\
\hline
\end{tabular}
\caption{\footnotesize Lowest p-values from various segments with corresponding frequencies, SNR, and Lorentzian fit parameters.}
\label{apptable0}
\end{table*}


{Before we apply this methodology to the power spectra reported in this work, we discuss some of its limitations and assumptions. It is assumed that in the absence of a QPO, the system has only Poisson noise. If the power spectra are Leahy normalized this would mean that $P_o = 2$. However, the LAXPC detector has dead time ($\tau_d \sim 42 \mu $ secs) which would change the expected Poisson noise level \citep{1995ApJ...449..930Z,yadav2016astrosat}. For frequencies less than $1/\tau_d$, the Poisson noise should still be frequency independent with values less than 2. However, it is important to check the effect of dead time on the probability estimation. We use the LAXPC simulator code \footnote{\url{http://astrosat-ssc.iucaa.in/proposal_preparation}} to generate ten simulated event files for the same spectra and count rate as the data and then used them to obtain simulated power spectra in the absence of any intrinsic variability.  The code incorporates a dead time of $42$ microseconds. We estimate $P_o$ using the average of the power in the frequency range under consideration, which is 400 to 1200\,Hz and then compute ${\mathrm p_N} (< P_{max})$ for the power spectra for different frequency bins. The results are plotted in Figure \ref{App0} (top panel), where we see that, as expected, ${\mathrm p_N} (< P_{max})$ is always less than 0.1. The formalism requires estimating the probability for different frequency bins, and this has not been taken into account in the overall probability computation. Since the maximum power for different frequency bins is correlated with each other, the probability estimates are not independent. If the search is undertaken for a relatively small number of frequency bins, the approach will still be reliable. There are also some effects which may lead to an overestimation of the probability, i.e. the results obtained may be conservative ones. Since $P_o$ is being estimated using the average of power at different frequencies, the presence of a QPO will increase $P_o$ above the Poisson level and hence would lead to an increase in ${\mathrm p_N} (< P_{max})$. Also, the formalism does not consider the possibility of more than one QPO in the power spectrum, and for such cases, the significance of the detection should be higher than the one estimated.}

{For all the power spectra reported in this work, we compute ${\mathrm p_N} (< P_{max})$ in the frequency range 400 -- 1200\,Hz for different frequency widths and plot the results in Figure \ref{App0}. The minimum ${\mathrm p_N} (< P_{max})$ obtained with the corresponding frequency and frequency width are listed in Table \ref{apptable0}, including the centroid frequency, width and SNR obtained from fitting as described in Table \ref{Table3}. The Table shows that for an SNR $>$ 3, the corresponding ${\mathrm p_N} (< P_{max})$ is less than 0.05, when one considers the number of frequency bins which are being searched for. Thus, the SNR  would correspond to a detection at $>$ 95\% confidence level if its value is larger than 3.}

\section*{Acknowledgments \& Data Availability }
{We thank the referee for their valuable comments}. We want to thank Dr. Thomas D. Russell for his insightful suggestions and discussions that helped us improve the result presented here. In this study, we utilized \textbf{LAXPC} and \textbf{SXT} data from {\em AstroSat}, provided by the Indian Space Science Data Centre (ISSDC). The research leading to these results has been funded by the Department of Space, Government of India, ISRO under grant no. DS\_2B-13013(2)/10/2020-Sec.2. Further, JC and RS acknowledge support from the Leverhulme Trust grant RPG-2023-240. We extend our gratitude to the AstroSat Science Support Cell (ASSC) for their assistance.  We also appreciate the efforts of the LAXPC and SXT Payload Operation Centers (POC) at TIFR, Mumbai. SC is thankful to IUCAA (Inter-University Centre for Astronomy and Astrophysics) for providing the periodic visit to carry out the research. Additionally, software from HEASARC was employed in this analysis. The LAXPC  and SXT archival data that has been used in this article can be found at {\em AstroSat} ISSDC website (\url{ https://astrobrowse.issdc.gov.in/astro\_archive/archive}).

\bibliography{Manuscript_1636}{}
\bibliographystyle{aasjournal}

\end{document}